# Joint encoding of “what” and “when” predictions through error-modulated plasticity in biologically-plausible spiking networks

Yohei Yamada[1], Zenas C. Chao[1]

1. International Research Center for Neurointelligence (WPI-IRCN), UTIAS, The University of Tokyo, Japan

Correspondence:
Zenas C. Chao (zenas.c.chao@gmail.com)
Yohei Yamada (yamada.yohei@mail.u-tokyo.ac.jp)

## Abstract

The brain anticipates future events using internal models that specify not only what will occur, but also when it will occur and with what probability. We refer to this joint specification of identity, timing, and likelihood as a *complete prediction object*. Existing computational models typically capture identity and timing separately, omit probability as an explicit representational dimension, or rely on biologically implausible global learning rules. Here we show that a single population of spiking neurons can acquire and flexibly maintain a complete prediction object through biologically grounded learning. We implemented a heterogeneous Izhikevich spiking reservoir with multiplexed readouts trained by an error-modulated, attention-gated three-factor Hebbian rule, and tested it on a task that independently manipulates event identity, latency, and probability. The network develops time-locked anticipatory activity whose amplitude scales with outcome probability and rapidly adapts when timing or probability statistics change. Identity and timing components self-organize into near-orthogonal readout subspaces within a shared neural population, demonstrating that multidimensional predictive structure can emerge without anatomical modularization or global error broadcast. Compared with least-squares-based approaches, local gated plasticity enables stable recalibration under nonstationary conditions. These results suggest that cortical mixed-selective populations, coupled with neuromodulator-gated synaptic plasticity, may be sufficient to jointly encode and update identity, timing, and probability within a single recurrent circuit. Flexible predictive cognition may therefore arise from generic population dynamics shaped by local learning rules rather than from specialized predictive modules.

## Introduction

Prediction is a defining property of adaptive neural systems. Across sensory, motor, and cognitive domains, neural activity often precedes external events, reflecting internal models that anticipate future states of the environment. Imagine standing in a garden and hearing a rustle in the bushes. From experience, you may expect that a bird will fly out quickly or that a cat will emerge more slowly, and you assign different likelihoods to these possibilities depending on context. Even before anything appears, the brain has formed a structured expectation about what will happen, when it will happen, and how likely it is. We refer to this joint specification of identity, timing, and likelihood as a *complete prediction object*.

This view implies that prediction is inherently multidimensional. To anticipate an event, the brain must represent its identity, its temporal structure, and its probability. Empirically, each of these components has been studied extensively. Cortical populations encode expected stimulus identity prior to onset (Summerfield et al., 2006; Kok et al., 2012; Mante et al., 2013; Bell et al., 2016; Meyer and Olson, 2011), parietal and frontal circuits represent elapsed time and hazard-like temporal structure (Cui et al., 2009; Janssen and Shadlen, 2005; Jazayeri and Shadlen, 2015; Nobre and van Ede, 2018), and probability modulates anticipatory magnitude and decision dynamics (Hangya and Kepecs, 2015; Shadlen and Newsome, 2001; Yang and Shadlen, 2007). Identity prediction has also been linked to oscillatory signatures, such as alpha and beta activity before stimulus onset (Arnal et al., 2011; Sedley et al., 2016; Bastos et al., 2020; Chao et al., 2022; Bauer et al., 2014), whereas temporal prediction often manifests as ramping activity that reflects the evolving hazard function over time (Mauk and Buonomano, 2004; Parker et al., 2014; Huang and Chao, 2025).

However, most of this literature examines these dimensions in relative isolation. Although a few studies have begun to bridge identity and timing within a common framework (Auksztulewicz et al., 2025, 2018; Bianco et al., 2024; Cappotto et al., 2023; Lau and Nguyen, 2015; Nara et al., 2021), the prevailing approach remains modular: identity, timing, and probability are typically treated as separable research domains. Moreover, probability is often formalized as a gain factor on identity or timing signals, rather than as an explicit representational axis. Thus, while the empirical components of prediction are well characterized, the computational question remains unresolved: how are identity, timing, and probability jointly learned and integrated within a biologically plausible circuit?

Computational accounts provide partial answers. Hierarchical predictive coding models formalize identity prediction through interactions between prediction and prediction error across cortical layers (Friston, 2010; Whittington and Bogacz, 2017), and recurrent loops can generate predictive feedback signals (Wacongne et al., 2012; Yaron et al., 2025). For temporal prediction, recurrent dynamical systems, including reservoir and FORCE-based approaches, show that high-dimensional population activity can approximate complex time-dependent functions (Laje and Buonomano, 2013; Nicola and Clopath, 2017; Sussillo and Abbott, 2009). Probabilistic inference has likewise been framed in terms of distributed population codes (Fiser et al., 2010; Ma et al., 2006). Yet many of these models rely on globally coordinated error signals or backpropagation-like updates with unclear circuit-level implementations (Bellec et al., 2020; Lillicrap et al., 2020), and few explain how a single neural population could jointly encode identity, timing, and probability while flexibly adapting to changing environmental statistics.

Here we ask whether a single recurrent spiking network, endowed with heterogeneous intrinsic dynamics and local neuromodulator-gated plasticity, is sufficient to learn the complete prediction object. Cortical circuits exhibit diverse intrinsic timescales and mixed selectivity that enable high-dimensional population dynamics (Buonomano and Maass, 2009; Murray et al., 2014; Rigotti et al., 2013), and synaptic plasticity is implemented locally under modulatory control (Brzosko et al., 2019; Frémaux and Gerstner, 2016; Seol et al., 2007). We hypothesize that prediction need not rely on persistent activity within specialized loops. Instead, it may emerge as a structured readout of recurrent dynamics, with prediction error acting primarily as a modulatory signal that gates local plasticity.

To evaluate this hypothesis, we operationalize the complete prediction object within a controlled cue-driven paradigm in which identity, timing, and probability are jointly specified and independently manipulated. This task allows us to test whether a single spiking reservoir with local, gated plasticity can (i) acquire calibrated identity and time-locked predictions, (ii) scale anticipatory magnitude with probability, and (iii) rapidly recalibrate when environmental statistics change. We then examine how these functions are expressed in weight space and population geometry, asking whether identity and timing emerge as separable yet overlapping subspaces within a shared neural population. By comparing local gated plasticity with global learning rules, we aim to clarify both the computational sufficiency and biological plausibility of single-population solutions to multidimensional prediction.

## Results

### Operationalizing the prediction object: Multi-Event Expectation Task

To test whether a single circuit can acquire the complete prediction object (Figure 1A), we designed a compact cue-driven paradigm that jointly manipulates identity, timing, and probability within each condition. We term this the Multi-Event Expectation Task (MEET). The task formalizes the three components of prediction introduced in the Introduction: which event will occur ("what"), when it will occur ("when"), and how likely it is ("probability").

A brief cue maps onto a structured prediction space composed of two discrete identities (Channel A or B) and a channel-specific latency distribution (Figure 1B). Within a given block, each channel is associated with a fixed interstimulus interval (ISI), while the realized channel on each trial is sampled according to a block-defined probability p(A). For example, one regime places Channel A at +30 ms post-cue offset with p(A)=80%, and Channel B at +90 ms with p(B)=20%. In this setting, the correct prediction object is not simply a label or a time point, but a probability-weighted, time-resolved expectation.

Trials are organized into 100-trial blocks with stationary statistics, allowing the model to learn stable prediction objects (Figure 1C). To test flexibility, block parameters are then switched abruptly, altering timing, probability, or both (Figure 1D). During evaluation, learning is disabled and only the cue is presented, so that anticipatory activity reflects the internally constructed prediction object rather than immediate sensory drive. Performance is quantified as the deviation between the predicted spatiotemporal trace and the ground-truth probability-weighted timing profile. This task therefore provides a minimal yet structured environment in which identity, timing, and probability must be jointly represented and rapidly recalibrated when environmental statistics change.

### Model design and architecture

We asked whether a single recurrent spiking population, equipped with local plasticity, can acquire and flexibly update the complete prediction object. The model consists of a reservoir of spiking neurons that receives three inputs during training: a Cue, stimulus A, and stimulus B. The recurrent population transforms these inputs into high-dimensional dynamics, and learning occurs only in the readout weights that generate predictions for Channel A and Channel B (Figure 2A).

If learning is successful, the Cue alone should be sufficient to evoke a structured prediction object. For example, under a condition in which stimulus A occurs with 80% probability at an earlier latency and stimulus B with 20% probability at a later latency, the Cue should trigger anticipatory activity that reflects both the probability and timing of the expected events (Figure 2B). In other words, the network should produce a probability-weighted, time-resolved prediction before any stimulus appears.

Figure 2C and Table 1 summarize the core design features and their implementation details, which are described below.

**_(1)_** **Fixed-weight heterogeneous Izhikevich reservoir** ***(Table 1A):*** The recurrent backbone is a fixed 1,000-neuron Izhikevich network (approximately 80% excitatory and 20% inhibitory) with heterogeneous intrinsic parameters (Izhikevich, 2004, 2003). Recurrent weights remain unchanged throughout training. The reservoir is not trained to memorize specific intervals. Instead, it provides a high-dimensional dynamical substrate. A brief cue evokes reproducible trajectories that unfold over tens to hundreds of milliseconds, spanning the temporal regime probed by MEET. Sparse, moderately strong recurrence places the network in a regime of rich but stable dynamics, such that trajectories are linearly readable without global error broadcast or backpropagation (Buonomano and Maass, 2009; Jaeger, 2001; Maass et al., 2002). Synaptic currents are modeled with double-exponential kernels and integrated at 1 ms resolution, matching the 10–100 ms latencies required for temporal prediction.

***(2) Multiplexed "what" and "when" from a shared population (Table 1B):*** Both identity and timing are decoded from the same subset of reservoir neurons. Timing prediction $z_{when}(t)$ is read out from the instantaneous filtered synaptic state $r(t)$, whereas identity prediction $z_{what}$ is read out from the post-cue average state $\bar{r}$. This architecture implements multiplexing: functional separation with anatomical overlap. The same mixed-selective neurons contribute to both predictions through distinct weight matrices $\Phi_{when}$ and $\Phi_{what}$. Thus, "what" and "when" are factorized in weight space rather than segregated into separate modules. Such shared-but-separable population geometry is consistent with cortical mixed selectivity and flexible linear readout mechanisms (Bell et al., 2016; Fusi et al., 2016; Mante et al., 2013; Rigotti et al., 2013).

***(3) Online timing learning versus offline identity consolidation (Tables 1C–D):*** The "when" pathway solves a temporal credit-assignment problem. It is updated online within each trial using a gated three-factor local Hebbian rule that combines presynaptic activity, signed timing error, and a phase-specific attention gate $G(t)$. Millisecond-level updates allow the model to align predicted activity precisely with the true latency window. In contrast, identity is trial-constant. The correct channel label does not depend on within-trial time. The “what” pathway is therefore updated once per trial using an ungated two-factor Hebbian rule based on the post-cue average state $\bar{r}$. This reduces moment-to-moment noise and supports stable calibration of channel probabilities across trials. This separation between fast, phasic timing updates and slower identity consolidation mirrors motifs observed in neuromodulated circuits while maintaining strictly local synaptic learning (Brzosko et al., 2019; Doya, 2002; Frémaux and Gerstner, 2016; Raymond and Medina, 2018; Yagishita et al., 2014).

***(4) Attention gate for temporal credit assignment (Table 1E):*** Timing updates are modulated by an attention gate $G(t)$ operating in two phases after cue offset. Before any teacher signal is detected, the gate is permissive across channels, allowing the model to learn a calibrated “predict-zero” baseline. When a teacher pulse occurs, the gate becomes selective: plasticity is enabled only for the detected channel and suppressed for the other until trial end. Importantly, the gate reacts only to the observed teacher pulse on that trial and does not access hidden ground truth. Functionally, this mechanism restricts synaptic credit to the relevant channel and latency window, preventing cross-talk and enabling rapid recalibration when the prediction object changes. Such gated plasticity resembles attention-dependent amplification and transient neuromodulatory learning windows in cortex (Brzosko et al., 2019; Buschman and Kastner, 2015; Frémaux and Gerstner, 2016; Reynolds and Heeger, 2009; Seol et al., 2007).

***(5) Stabilizing timing feedback (Table 1F):*** During cue-only evaluation, the model’s own timing prediction is fed back into the reservoir through a fixed, sign-constrained projection with modest gain: $I_{feedback}(t) = E^{\top} z_{when}(t)$. This self-generated feedback stabilizes phase alignment of trajectories and reduces drift without leaking teacher information. Adaptation occurs exclusively through readout plasticity and closed-loop modulation, while the recurrent backbone remains fixed (Buschman and Kastner, 2015; Laje and Buonomano, 2013; Reynolds and Heeger, 2009).

***(6) Integration into a complete prediction object (Table 1G):*** Finally, identity and timing outputs are combined multiplicatively: $\hat{y}(t) = z_{what} \times z_{when}(t)$. Identity specifies which channel is expected and with what confidence. Timing specifies when that expectation should manifest. Their product yields a single probability-weighted, time-resolved prediction trace, corresponding to the complete prediction object. Because identity and timing remain separable in weight space, errors can be attributed to either dimension, allowing rapid updating under nonstationary statistics without architectural modification. Multiplicative gain modulation is consistent with canonical cortical mechanisms for context-dependent processing (Carandini and Heeger, 2011; Chance et al., 2002; Reynolds and Heeger, 2009; Salinas and Abbott, 1996).

**Robust event prediction and probability tuning**

Using our proposed model, we first tested whether the circuit can generate accurate event predictions in both identity and timing, and whether it encodes event probability when two events with distinct identities and latencies occur with different likelihoods. We fixed the latencies at 30 ms for A and 90 ms for B, and varied the ”what” probability p(A). Under cue-only evaluation, prediction amplitudes scaled systematically with p(A): when p(A) was high, A-selective activity dominated the 30–60 ms window; when p(A) was low, B-selective activity dominated the 90–120 ms window (Figure 3A). Window-averaged responses varied monotonically with p(A) and exhibited low trial-to-trial dispersion (Figure 3B), indicating robust generalization across the full range of event probabilities.

Quantitatively, window-averaged A predictions (30–60 ms) tracked p(A) closely (Pearson $r = 0.9779, p = 2.01 \times 10^{-7}$; OLS $y = 0.8990 \times p(A) - 0.0685$; $R^2$=0.8113; slope $t = 14.0360$, $p = 2.01 \times 10^{-7}$). B predictions (90–120 ms) anticorrelated with p(A) and matched the complementary mapping to p(B)=1-p(A) (Pearson $r = -0.9891$, $p = 8.62 \times 10^{-9}$; OLS $y = -1.0147 \times p(A) + 0.9681$; $R^2 = 0.9614$; slope $t = -20.1181$, $p = 8.62 \times 10^{-9}$). Monotonic trend tests confirmed consistent ordering across all 11 probability levels (Spearman $\rho_A$=0.9909, $\rho_B$=-0.9909; both $p = 3.76 \times 10^{-9}$). This amplitude-probability tuning echoes probability-matching and Bayesian-like weighting in cortical decision circuits, yet here it emerges without explicit likelihood computation at test time.

**Rapid learning of the complete prediction object**

Next, we examined the learning dynamics within a block and asked whether the circuit can establish the prediction object under different timing combinations. As a concrete example, we focused on the condition with ISI [30, 90] ms and p(A) = 50%, one of the probability settings shown in Figure 3, and tracked how predictions emerged across 100 trials. Figure 4A illustrates the block structure used for this analysis. Figure 4B shows the evolution of the identity (“what”) prediction across trials, demonstrating gradual calibration toward the true 50–50 probability. Figure 4C presents the temporal (“when”) predictions as trial-by-trial heat maps, where activity progressively concentrates around the correct latency windows (white markers). Figure 4D shows the fused what × when output, confirming that identity and timing are jointly expressed in a single, probability-weighted temporal trace. Consistent with these qualitative patterns, the root-mean-squared error decreased rapidly over trials (Figure 4E), indicating efficient learning of the complete prediction object within a single block.

The same signatures of calibrated probability estimates, time-locked temporal responses, and low prediction error were observed under other timing and probability conditions: ISI [60, 60] ms with p(A)=50% (Figure 4F), ISI [90, 30] ms with p(A)=50% (Figure 4G), ISI [30, 90] ms with p(A)=80% (Figure 4H), and ISI [30, 90] ms with p(A)=100% (Figure 4I). Notably, when both channels shared the same probability and latency (p(A)=50% with ISI [60, 60] ms; Figure 4F), discrimination information was minimal, so RMSE decreased more slowly and to a higher asymptote; even so, the model correctly placed its anticipated timing at +60 ms and continued to improve across trials. Quantitatively, using the mean RMSE over the first 10 trials as a universal baseline, end-of-block (trial 100) errors were significantly lower in all five representative conditions: baseline means $0.1511 - 0.2016(\pm 0.0103 - 0.0294)$, trial-100 RMSE $0.0667 - 0.1503$, giving $\Delta RMSE = 0.0513 - 0.0975$ with strong statistical support ($t = 8.42 - 14.91$; all $p \leq 7.34 \times 10^{-6}$). In sum, under stationary conditions, the model rapidly and reliably encoded the complete prediction object.

**Rapid adaptation to different prediction objects**

We next asked whether the model can recalibrate the prediction object when task statistics change. We used two-block sequences with an abrupt switch after trial 100, with architecture and hyperparameters held fixed. We tested two manipulations: timing reversals (ISI [30, 90] →[90, 30] ms) at p(A)=50% (Figure 5A), 80% (Figure 5B), and 100% (Figure 5C), and probability flips (100%→0% (Figure 5D), 80%→20% (Figure 5E), 60%→40% (Figure 5F) at ISI [30, 90] ms). Under these switches, the circuit recalibrated rapidly without overshoot or drift. Identity estimates tracked the new probabilities with minimal transients, temporal precision was preserved despite interval changes, and the fused output maintained spatiotemporal fidelity. RMSE showed a brief, expected bump at the block boundary followed by rapid reconvergence. Quantitatively, across six adaptation schedules, end-of-block errors were consistently below the within-condition baseline (first 10 trials). By the end of Block 1, RMSE ranged 0.0667–0.0921 (ΔRMSE = 0.0826–0.1045; t = 8.42–10.45; all $p \leq 7.34 \times 10^{-6}$). By the end of Block 2, RMSE ranged 0.0513–0.0917 (ΔRMSE = 0.0791–0.0998; t = 9.06–17.64; all $p \leq 4.04 \times 10^{-6}$).

Mechanistically, this rapid recalibration arises from three complementary design features. First, timing updates are performed online and gated by channel-specific attention signals, so synaptic changes are restricted to the active channel and its latency window, preventing interference across dimensions. Second, identity learning is consolidated offline at trial end, which stabilizes probability estimates while remaining sensitive to block-wise shifts. Third, closed-loop timing feedback stabilizes the reservoir trajectory around the newly predicted latency manifold, reducing drift during cue-only evaluation. Together, these local and phase-specific adjustments allow synaptic credit to be reassigned in place, enabling the model to update the contents of the prediction object without modifying the recurrent backbone.

**Flexible multi-object adaptation**

We further tested whether rapid recalibration persists across multiple successive switches. To test this, we ran three four-block schedules with weights carried forward, architecture and hyperparameters held fixed. Schedule 1 alternated probabilities while introducing a late timing reversal (p(A)=80%→20%→80%→20%; ISI [30,90]→[30,90]→[90,30]→[90,30] ms; Figure 6A). Schedule 2 paired probabilities by halves while flipping timing every block (p(A)=80%→80%→20%→20%; ISI [30,90]→[90,30]→[30,90]→[90,30] ms; Figure 6B). Schedule 3 fixed identity at p(A)=100% but swept A's latency systematically ([30,120]→[60,90]→[90,60]→[120,30] ms; Figure 6C).

Under Schedule 1 (Figure 6A), identity estimates tracked "what" probabilities with only brief boundary transients, predicted latencies remained time-locked to the correct windows after each switch, and the fused what × when × probability output aligned with ground truth across all four blocks. Error trajectories showed brief peaks at each transition followed by rapid recovery: relative to the within-session baseline (trials 1–10; 0.1642 ± 0.0240), end-of-block RMSE at trials 100, 200, 300, and 400 was 0.0667, 0.0742, 0.0889, and 0.0596, corresponding to improvements of 0.0975, 0.0900, 0.0753, and 0.1046 ($t$ = 12.18, 11.25, 9.41, 13.07; all $p \leq 3.4\times10^{-7}$). Under Schedule 2 (Figure 6B), the circuit again recalibrated rapidly despite alternating the timing every block. Using the same baseline (0.1642 ± 0.0240), end-of-block RMSE was 0.0667, 0.0917, 0.0580, and 0.0654, yielding improvements of 0.0975, 0.0725, 0.1062, and 0.0988 ($t$ = 12.18, 9.06, 13.27, 12.35; all $p \leq 4.0\times10^{-6}$). Under Schedule 3 (Figure 6C), with p(A) fixed at 100% and progressively shifted latencies across blocks, recalibration remained rapid, though Block 2 imposed a larger timing change. Relative to its baseline (0.1511 ± 0.0294), end-of-block RMSE was 0.0684, 0.1064, 0.0795, and 0.1080, for improvements of 0.0826, 0.0447, 0.0716, and 0.0431 ($t$ = 8.42, 4.55, 7.29, 4.39; all $p \leq 0.001$).

**Model comparisons under multi-object adaptation**

We next compared our model and its variants with existing approaches using the same four-block schedule, in which probability alternated across blocks (p(A): 80%→20%→80%→20%) and latency changed at every block ([30,120]→[60,90]→[90,60]→[120,30] ms). The result from our model is shown in Figure 7A. We also evaluated FORCE (RLS) in an otherwise identical architecture, including the same gated "when" pathway, so that any difference isolates the effect of replacing local three-factor plasticity with a global, error-driven update (Figure 7B, denoted as "Gated FORCE"). Moreover, we included an offline variant that withholds real-time updates in the "when" pathway and adjusts weights only after trials, testing whether removing within-trial timing updates slows or degrades recalibration (Figure 7C, denoted as "Offline variant"). Finally, a single-readout baseline collapses "what" and "when" into one channel, eliminating spatial separation and probing whether failure to factorize identity and timing induces interference when the prediction object changes (Figure 7D, denoted as "Single-stream variant").

Our proposed model maintained low error and rapid post-switch recalibration. For "what" prediction, it tightly tracked ground-truth probabilities across every switch; For "when" prediction, it preserved crisp, time-locked responses at the correct windows; and for fused what × when prediction, it retained spatiotemporal fidelity throughout. For RMSE, relative to its baseline (0.1573 ± 0.0251), end-of-block values at trials 100, 200, 300, and 400 were 0.0935, 0.0751, 0.1209, and 0.0808, respectively (improvements 0.0638, 0.0822, 0.0363, 0.0765; $t$ = 7.63, 9.84, 4.35, 9.16; $p = 1.61\times10^{-5}$, $2.05\times10^{-6}$, $9.28\times10^{-4}$, $3.70\times10^{-6}$).

The Gated FORCE model fit the first stationary block but adjusted poorly at subsequent boundaries. From a baseline of 0.1761 ± 0.0144, Block 1 ended at 0.1015 ($\Delta = 0.0746$; $t = 15.59$; $p = 4.04\times10^{-8}$). There was no significant improvement by the end of Blocks 2–4, and endpoints were at or above baseline (0.1675, 0.2038, 0.1801; $\Delta$ = 0.0086, −0.0277, −0.0040; $t$ = 1.80, −5.79, −0.83; one-sided $p$ = 0.0531, ≈1.0, 0.7860). The Offline-variant model also improved across blocks: relative to its baseline (0.2301 ± 0.0467), end-of-block RMSE was 0.1229, 0.0619, 0.2236, 0.1032 ($\Delta$ = 0.1072, 0.1682, 0.0065, 0.1269; $t$ = 6.89, 10.81, 0.42, 8.16; $p = 3.57\times10^{-5}$, $9.32\times10^{-7}$, 0.3420, $9.47\times10^{-6}$), though it remained higher than the proposed model. The Single-stream variant model showed improvement but was noisier (high baseline variance 0.1648 ± 0.0216). End-of-block RMSEs were 0.1045, 0.0966, 0.1448, and 0.1074 ($\Delta$ = 0.0603, 0.0682, 0.0199, 0.0574; $t$ = 8.39, 9.48, 2.77, 7.98; $p = 7.57\times10^{-6}$, $2.78\times10^{-6}$, 0.0108, $1.13\times10^{-5}$).

Together, these comparisons indicate that rapid multi-object adaptation depends on three interacting elements: local synaptic credit assignment, millisecond-level online timing updates, and explicit factorization of identity and timing within shared readout populations. When any one of these components is removed, the system either adapts more slowly, becomes unstable across switches, or exhibits interference between dimensions. This pattern supports the idea that stable yet flexible encoding of the prediction object requires both functional separation and anatomically shared representations, coordinated through local, gated plasticity.

**Readout weight dynamics underlying prediction object encoding**

To test whether learning factorizes the prediction object into separable components, we analyzed the readout weights, the locus of plasticity and the interface to behavior, rather than transient reservoir activity. For each trial, we concatenated the weight vectors from the "what" and "when" streams for both A and B channels (300 readout neurons × 2 streams × 2 channels), yielding a 1,200-dimensional feature vector. Principal component analysis (PCA) was performed on weight vectors collected from single-block simulations spanning 20 task conditions (5 p(A) levels × 4 ISI conditions; 100 trials each) to define a reference subspace.

To assess adaptability under nonstationarity, we then ran a multi-block simulation in which the same 20 conditions were presented in pseudorandom order across three consecutive blocks (60 blocks total). Weight trajectories from this run were projected into the single-block reference space. Both the proposed model and the Gated FORCE model were subjected to the identical schedule, enabling direct comparison of representational geometry and adaptation (see Methods).

In this space, the proposed model organized along near-orthogonal axes. The first principal component (PC1) sorted conditions primarily by probability, whereas the second principal component (PC2) sorted them by latency (Figure 8A). In single-block runs, trajectories evolved smoothly from a compact origin toward condition-specific endpoints (Figure 8B). In multi-block sequences, trajectories consistently turned toward their corresponding single-block endpoints rather than drifting elsewhere (Figure 8C). When aggregated across repetitions, the latent positions aligned closely with independently obtained single-block references (Figure 8D). These results indicate that learning induces stable, statistics-dependent attractors in readout space, with probability and latency encoded along separable dimensions.

The Gated FORCE model showed a different pattern. In single-block simulations, it also formed well-separated representations with orderly trajectories (Figures 8E and 8F), and during the first block of multi-block runs it reached appropriate configurations (Figure 8G, Block 1). However, after subsequent switches, trajectories progressively collapsed toward a central region of weight space (Figure 8G), erasing separability and preventing accurate tracking of later task conditions. Distances between condition-specific representations remained large and failed to reconverge across blocks (Figure 8H). This pattern resembles catastrophic interference, where global updates disrupt previously established structure and undermine stable context-specific solutions.

**Structured plasticity supports rapid adaptation under nonstationarity**

The representational geometry described above was directly linked to learning performance. We quantified stability as the Euclidean distance between each multi-block weight vector and its condition-matched single-block endpoint (Figure 8I), computed across 60 condition-block pairs. The proposed model maintained consistently smaller distances throughout learning, indicating rapid convergence toward condition-specific attractors and stable alignment with the single-block reference geometry. In contrast, Gated FORCE exhibited larger and more persistent deviations, particularly after block switches, consistent with drift and partial collapse in weight space.

These geometric differences translated into behavioral error (Figure 8J). We compared root-mean-square error (RMSE) at trials 1, 10, and 100 within each block across 60 blocks using one-sided t-tests (Proposed Model < Gated FORCE). At trial 1, there was no advantage for the proposed model. It started slightly worse (0.2203 ± 0.0585) than Gated FORCE (0.1947 ± 0.0269; $t = 3.0565$, $p = 0.998615$, n.s.). By trial 10, however, the proposed model exhibited significantly lower error (0.1528 ± 0.0539 vs 0.1933 ± 0.0259; $t = -5.2111$, $p = 4.04\times10^{-7}$). By trial 100, the difference widened markedly (0.0811 ± 0.0191 vs 0.1788 ± 0.0228; $t = -25.1834$, $p < 10^{-15}$).

Thus, the same synaptic factorization that preserves separable “what” and “when” axes is accompanied by faster and more stable error reduction under nonstationarity. In the proposed model, local, attention-gated updates maintain condition-specific structure in readout space, enabling rapid convergence toward task-appropriate attractors after switches. In contrast, globally coupled least-squares updates reshape the entire readout subspace after each error, leading to representational drift and slower recovery.

To test whether limited adaptability in Gated FORCE could be attributed to insufficient forgetting of past statistics, we introduced a forgetting factor into the recursive least squares update and systematically varied it across a broad range (see Supplementary Method 1 and Supplementary Figure 1). Although moderate forgetting slightly increased short-term responsiveness, no parameter setting restored stable geometry or matched the performance of the proposed model. Varying the learning-rate related hyperparameters of Gated FORCE likewise did not improve adaptability (see Supplementary Method 2 and Supplementary Figure 2). Together, these control analyses suggest that the limitation reflects the global structure of the update rule, rather than a particular choice of hyperparameters.

# Discussion

We proposed that predictive signals need not reside in anatomically segregated modules, but can instead be instantiated in structured readout weights operating on a shared recurrent substrate. Specifically, we hypothesized that prediction errors act as local modulators of plasticity rather than global teaching signals, that "what" and "when" emerge as separable yet overlapping subspaces within mixed-selective populations, and that probability is expressed as graded gain on anticipatory activity. The present results provide a minimal demonstration of these principles. Within a fixed, heterogeneous reservoir, the complete prediction object is encoded in readout geometry that reorganizes under local, attention-gated plasticity to form near-orthogonal axes for identity and latency while scaling magnitude with likelihood. The model is not a reconstruction of cortical circuitry, but a proof of principle that a single recurrent population with biologically grounded local learning rules is sufficient to generate multiplexed predictive signals and recalibrate them under nonstationary statistics.

**Model principles and computational distinction**

The architecture assembles common physiological motifs into a minimal predictive configuration. A heterogeneous recurrent network supplies high-dimensional, state-dependent dynamics that are linearly readable (Buonomano and Maass, 2009; Jaeger, 2001; Maass et al., 2002). Adaptation is confined to the readout layer, demonstrating that flexible predictive structure can emerge from synaptic reweighting without recurrent rewiring. Although recurrent connectivity remains fixed, closed-loop feedback reshapes effective dynamics as readout weights evolve, altering attractor geometry through input modulation (Laje and Buonomano, 2013).

Timing plasticity follows a phase-specific three-factor rule consistent with neuromodulator-gated eligibility traces (Brzosko et al., 2019; Frémaux and Gerstner, 2016; Yagishita et al., 2014), while identity consolidates at the trial level. This separation enables selective credit assignment and addresses the stability–plasticity problem without global error broadcast. In contrast to global least-squares approaches such as FORCE/RLS, which couple all weights through a shared inverse-correlation state (Nicola and Clopath, 2017; Sussillo and Abbott, 2009), local three-factor updates modify only synapses implicated in the active channel and temporal window. This locality clarifies why adaptation remains rapid and drift-free under context switches.

These results align with cortical evidence that mixed-selective populations multiplex task variables and that downstream projections define functional axes (Fusi et al., 2016; Mante et al., 2013; Rigotti et al., 2013). High-dimensional codes expand the space of linearly accessible solutions and support flexible recombination under uncertainty (Murray et al., 2014).

**Biological relevance and testable hypotheses**

The framework provides a biologically grounded account of how predictive coding may be implemented within recurrent cortical circuits (Figure 9). In this formulation, prediction errors function primarily as modulators of plasticity rather than as dominant feedforward drivers of representation. Predictive activity itself propagates forward through the hierarchy, while transient error-related signals gate synaptic updates locally and adjust feedback gain. Thus, feedforward transmission carries a multiplexed mixture in which predictions dominate representation and errors regulate learning.

The primary hypotheses (Table 2) capture the essential computational principles. First, prediction errors operate as local modulatory signals that transiently gate synaptic plasticity and feedback stabilization. Second, "what" and "when" predictions form partially orthogonal subspaces embedded within overlapping neural manifolds, consistent with mixed selectivity observed in sensory and prefrontal cortices. Third, intrinsic dynamics shape predictive precision, such that manipulations of cortical variability, arousal, or gain should systematically alter timing accuracy and probability scaling.

The secondary hypotheses (Table 3) specify candidate biological mechanisms. Phase-specific bursts of acetylcholine, dopamine, or norepinephrine are predicted to gate timing plasticity within defined post-detection windows. Closed-loop feedback from readout projections, represented as $I_{\text{feedback}}$, should stabilize temporal manifolds and reduce latency jitter. Multiplicative fusion of identity and timing may correspond to cortical gain modulation integrating identity confidence with temporal precision. Finally, a two-phase baseline process may establish a "predict-zero" state before event detection, providing a dynamic reference against which prediction and error are computed.

Each hypothesis yields concrete experimental tests. In humans, EEG combined with pharmacological manipulation of cholinergic or dopaminergic tone should selectively alter timing recalibration while sparing identity encoding. In rodents, Neuropixels recordings paired with optogenetic perturbation during specific post-detection windows should disrupt latency updating without globally degrading representation. Population analyses using demixed PCA or subspace alignment methods can test whether identity and timing indeed occupy partially orthogonal manifolds embedded within shared neural populations.

Together, these predictions link local synaptic gating, mixed selectivity, and feedback stabilization into a unified neurophysiological account of how cortical circuits might construct and update a complete prediction object.

**Limitations and future directions**

The present model abstracts away from several biological complexities. The two-alternative design simplifies naturalistic settings with concurrent events, omissions, and graded hazard structures. The multiplicative fusion of identity and timing likely underestimates nonlinear cortical integration. Moreover, our analyses emphasize readout geometry; future work should characterize how reservoir state manifolds reorganize during adaptation and how upstream dynamics contribute to predictive encoding.

Extensions to hierarchical or stacked reservoirs may better capture cross-level prediction–error interactions observed in cortical hierarchies (Chao et al., 2022, 2018). Causal perturbation of neuromodulatory systems during defined temporal windows would provide direct tests of the proposed gating mechanism. Broadly, these results suggest that heterogeneity and mixed selectivity are enabling features of predictive computation: a generic recurrent substrate, properly gated by local plasticity, is sufficient to bind what and when while remaining flexible under changing statistics.

## Methods

### Task paradigm

We implemented a dual-channel temporal prediction task (Multi-Event Expectation Task, MEET) that requires simultaneous encoding of channel identity ("what") and temporal latency ("when") under probabilistic conditions. Each trial contains two mutually exclusive channel identities (A or B), each associated with a discrete latency. Trials are arranged in 100-trial blocks with fixed parameters; at block boundaries, probabilities and/or latencies switch abruptly to induce nonstationarity while the network's internal state is preserved across trials to maintain realistic temporal dynamics.

After a jittered inter-trial interval (ITI; 0–200 ms), each trial begins with a 100 ms baseline, followed by a 30 ms cue. After cue offset, exactly one channel becomes active (i.e. 'teacher signal') at its channel-specific interstimulus interval (ISI). During training, typically 30–120 ms post-cue offset, Channel A signal appears at a discrete delay with probability p(A), and Channel B appears at its delay with p(B) = 1 − p(A). Each activation is a 30 ms window, producing distinct spatiotemporal patterns that require concurrent learning of identity and timing. Following each individual training trial run, test trial run is conducted online, but being presented with the cue alone (teacher signal withheld; learning disabled). This train-then-test cycle repeats on every trial throughout each 100-trial block and continues across consecutive blocks with the preset p(A) and ISI settings.

This structure poses two coupled challenges: selecting which channel will occur on each trial (stochastic "what") and predicting when it will occur (deterministic "when" within a block). Within a block, ISIs are fixed (e.g., A = 30 ms, B = 90 ms), while the realized channel on each trial is sampled pseudo-randomly from p(A), p(B)=1-p(A) using a fixed seed for each task condition. The combination of fixed temporal structure with probabilistic identity yields a compact yet rich environment in which models must extract statistical regularities while maintaining precise temporal predictions.

### Training and testing protocol

Each trial consisted of a training phase (cue + teacher signal, learning enabled) immediately followed by a testing phase (cue only, learning disabled). This one-trial mini-batch structure ensured that readout weights were updated after each individual trial.

***Training Phase:*** A 30 ms cue was presented, followed by a teacher signal indicating both the identity ("what") and timing ("when") of the expected event. The active channel (A or B) for each trial was sampled according to the block's ground-truth probability:

1. Channel A was selected with probability p(A),
2. Channel B with probability 1 - p(A).

The selected channel received a 30 ms rectangular pulse delivered at the channel-specific latency (e.g., 30 ms or 90 ms after cue offset), providing ground-truth temporal information.

Readout weights were updated separately for timing and identity decoders:

1. $\Phi_{\text{when}}$ was updated within the trial using attention-gated three-factor Hebbian plasticity.
2. $\Phi_{\text{what}}$ was updated at the end of the trial, based on averaged post-cue activity.

For the Gated FORCE variant, recursive least squares (RLS) replaced local Hebbian updates during the timing phase.

***Integration Mechanism:*** At test time, the model generates a complete prediction object $\mathrm{z}_{\text{prediction object}}$ , which represents the full spatiotemporal probability structure of the expected event – i.e., both *which* channel is predicted to be active and *when* it is expected to occur. This object is constructed by multiplicatively combining two components:

1. Identity prediction $\mathrm{z}_{\text{what}}$, a static per-trial probability estimate of the active channel, and
2. Timing prediction $\mathrm{z}_{\text{when}}(t)$, a time-resolved estimate of event latency.

Mechanistically, the fusion is computed at each time step $t$ as:

$$\mathrm{z}_{\text{prediction object}}(t) = \mathrm{z}_{\text{what}} \times \mathrm{z}_{\text{when}}(t).$$

Here, $\mathrm{z}_{\text{what}}$ acts as a channel selector, while $\mathrm{z}_{\text{when}}(t)$ provides the temporal profile. This multiplicative interaction is functionally analogous to an AND gate, where identity gates the relevant output channel and timing determines the precise latency. The resulting fused representation $\mathrm{z}_{\text{prediction object}}$ thus encapsulates the model's complete predictive, probabilistic belief over identity and time.

***Testing Phase:*** During testing, only the cue was presented. No teacher signal was delivered, and learning was disabled. The model generated predictions of the fused what × when × probability output (i.e., the complete prediction object) based on the learned decoders. Mini-batch size was fixed at 1 (one trial per update).

### Heterogenous Izhikevich spiking reservoir

All models shared the same neural substrate: a 1,000-neuron Izhikevich spiking reservoir with identical connectivity and parameter distributions (heterogeneous a, b, c, d; mixture of regular-spiking, fast-spiking, and bursting cells), as governed by:

$$\frac{dv_i}{dt} = 0.04{v_i}^2 + 5v_i + 140 - u_i + I_i(t)$$

$$\frac{du_i}{dt} = a_i(b_i v_i - u_i)$$

with after-spike reset: if $v_i \geq 30mV$, then $v_i \leftarrow c_i;\ u_i \leftarrow u_i + d_i$. Parameters were heterogeneously distributed to provide temporal basis diversity:

Excitatory neurons (80%):

$$a_i \sim U(0.01 \pm 0.02)$$

$$b_i \sim U(0.10 \pm 0.20)$$

$$c_i \sim U(-55 \pm 20)$$

$$d_i \sim U(2 \pm 4)$$

Inhibitory neurons (20%):

$$a_i \sim U(0.05 \pm 0.10)$$

$$b_i \sim U(0.15 \pm 0.10)$$

$$c_i \sim U(-65 \pm 5)$$

$$d_i \sim U(1 \pm 2)$$

where $U(\mu \pm \sigma)$ denotes uniform distribution over $[\mu - \sigma, \mu + \sigma]$. Readouts comprised 300 neurons (30% of the reservoir), ensuring matched capacity across methods.

### Gated local Hebbian model

A dual-pathway readout (Figure 2B) uses the same anatomical pool of neurons to encode identity ("what") and timing ("when") with distinct weight sets $\Phi_{\text{what}}$ and $\Phi_{\text{when}}$. The reservoir receives the cue input $I_{\text{cue}}(t)$, recurrent drive $\Omega$, a small background bias $I_{\text{bias}}(t)$, and a Dale's-law sign-constrained, sparse timing feedback $I_{\text{feedback}}(t) = E^{\top} z_{\text{when}}(t)$.

***Readout mappings (predictions):*** Two linear readouts act on the same sparse subset of reservoir neurons (multiplexing): a timing readout for "when" and an identity readout for "what." Let $\Phi_{\text{when}}, \Phi_{\text{what}} \in R^{N\times O}$ (with O=2 channels) and $\phi_{\text{mask}} \in \{0,1\}^{N\times O}$ be a fixed binary mask selecting the shared readout pool:

- Timing (online): $z_{\text{when}}(t) = \sigma\left(\boldsymbol{r}(t)^{\top}(\Phi_{\text{when}} \odot \phi_{\text{mask}})\right) \in (0,1)^{O}$,
- Identity (offline): $z_{\text{what}} = \sigma\left(\overline{\boldsymbol{r}}^{\top}(\Phi_{\text{what}} \odot \phi_{\text{mask}})\right) \in (0,1)^{O}$.

Here $\odot$ denotes the Hadamard (element-wise) product; $\sigma$ denotes a sigmoid function. Learning occurs only at $\Phi_{\text{when}}$ and $\Phi_{\text{what}}$; the reservoir $\Omega$ and the Dale-constrained feedback $E$ are pseudo-randomly initialized and fixed throughout.

***Online timing ("when")– gated three-factor Hebbian:*** Let $y_{\text{when}}(t) \in \{0,1\}^{O}$ be the teacher timing series (1 inside the true latency window of the active channel; 0 otherwise), and $G(t) \in \{0,1\}^{O}$ the attention gate (defined below). The within-trial update is

$$\Delta\Phi_{\text{when}}(t) = \eta_{\text{when}} \times \boldsymbol{r}(t) \times \left([y_{\text{when}}(t) - z_{\text{when}}(t)] \odot G(t)\right)^{\top} \odot \phi_{\text{mask}}.$$

This is a strictly local three-factor rule: pre-synaptic $\boldsymbol{r}(t)$ × error $[y_{\text{when}} - z_{\text{when}}]$ × gate $G(t)$, applied element-wise on the masked weights.

***Attention gate $G(t)$ (applies only to "when"):*** Let $t_{\text{detect}}$ be the first time any channel's teacher series turns 1 following cue offset (event onset):

- Pre-detection ($t < t_{\text{detect}}$): $G_i(t) = 1$ for all channels (learn calibrated predict-zero baselines).
- Post-detection ($t \geq t_{\text{detect}}$): $G_i(t) = 1$ only for the active channel; $G_{j\neq i^*}(t) = 0$ (restrict credit to the correct channel and its latency window).

The identity pathway effectively uses $G \equiv 1$ because it updates offline from $\bar{r}$ against a trial-constant label. See Supplementary Table 1 for further details on the implementation of the Gated Local Hebbian model.

***Offline identity ("what")– ungated two-factor Hebbian (trial end):*** Let $y \in \{0,1\}^{O}$ be the one-hot identity label. Using time average of $r(t)$ from cue offset to trial end $\bar{r}$:

$$\Delta\Phi_{\text{what}} = \eta_{\text{what}} \times \overline{\boldsymbol{r}} \times (y_{\text{what}} - z_{\text{what}})^{\top} \odot \phi_{\text{mask}},$$

where $\bar{r}$ is the post-cue double-exponential synaptic filtered average across time of the trial. This is a strictly local two-factor rule, and no gate is used for identity (target is trial-constant).

***State used by the readouts (double-exponential synaptic filtering):*** Let $\boldsymbol{r}(t) \in R^N$ be the filtered reservoir state vector of synaptic activity (a firing-rate proxy obtained by double-exponential filtering of spikes). Spikes are passed through a fast-rise, slower-decay synapse to produce the continuous "filtered state" consumed by the readouts:

$$r(t) = r(t-1)\left(1 - \frac{\Delta t}{\tau_r}\right) + h_r(t-1)\Delta t,$$

$$h_r(t) = h_r(t-1)\left(1 - \frac{\Delta t}{\tau_d}\right) + s(t)\,\kappa,$$

with $\Delta t = 1\,\mathrm{ms}$, rise constant $\tau_r$, decay constant $\tau_d$, and scale $\kappa$.

**Gated FORCE**

To isolate the effect of the learning rule, we implemented a Gated FORCE model that used the same network architecture and gating mechanism as the proposed model but replaced local Hebbian updates with FORCE-based recursive least squares (RLS) learning for the readouts. Weights were updated according to the standard RLS step $\phi_t = \phi_{t-1} - P_{t-1} r_t e_t^{\top}$ with the Sherman-Morrison update for the inverse correlation matrix. For the when pathway, the same gate as in the proposed model was applied by using a gated error $e_t = \left(y_{\mathrm{when}}(t) - \hat{x}_{\mathrm{when}}(t)\right) \odot G(t)$ at each millisecond timestep. The what pathway received a single trial-end update using the post-cue average activity $\bar{r}$ without gating. To increase responsiveness at block boundaries, an optional forgetting factor $0 < \lambda_f \leq 1$ was incorporated into the P matrix update:

$$P_t = \lambda_f^{-1}\left[P_{t-1} - \frac{P_{t-1} r_t r_t^{\top} P_{t-1}}{\lambda_f + r_t^{\top} P_{t-1} r_t}\right] \; (with\, P_0 = \alpha^{-1} I)$$

and $\lambda_f$ was swept across a broad range. Values of $\lambda_f < 1$ accelerated early adjustments but also increased variance and did not match the drift-free, rapid recalibration achieved by the proposed model across multi-block schedules. Results shown in this paper use $\lambda_f = 1$, but see Supplementary Figure 1 for results testing $\lambda_f \in \{1.0, 0.999999, 0.99999, 0.9999, 0.999\}$.

**Offline Hebbian**

Identical architecture to the proposed model, but timing updates are deferred to the end of each trial, with no within-trial ("online") updates to $\Phi_{\mathrm{when}}$. At trial end, when weights use a two-factor supervised Hebbian step (pre × error, no gate):

$$\Delta\Phi_{\mathrm{when}} \propto \eta_{\mathrm{when}}\left(\sum_{t \in \mathrm{postcue\ offset}} r(t) \times \left[\,y_{\mathrm{when}}(t) \;-\; z_{\mathrm{when}}(t)\,\right]^{\top}\right) \odot \phi_{\mathrm{mask}}.$$

Equivalently, one can write $\sum_t r(t)$ and $\sum_t [y_{\text{when}}(t) - z_{\text{when}}(t)]$ as trial-averaged eligibility and error terms. The what pathway is unchanged from the main model and is also two-factor at trial end:

$$\Delta\Phi_{\text{what}} \propto \eta_{\text{what}} \times \bar{r} \times [\, y - z_{\text{what}} \,]^{\top} \odot \phi_{\text{mask}},$$

with $\bar{r}$ the post-cue average across time of the trial. This baseline therefore removes the phase-specific attention gating and millisecond-level credit assignment used by the online when learner, directly testing whether purely offline two-factor updates can support rapid, drift-free recalibration under switches (anticipating slower and noisier adaptation).

**Single Hebbian**

A simplified control that collapses "what" and "when" into a single output stream (no explicit factorization into separate pathways). This baseline tests whether the decomposition of identity and timing is necessary for stable, accurate prediction under the task's probabilistic identity and fixed-latency structure.

All models were initialized with identical pseudo-random seeds (reservoir weights and neuron parameters, readout initializations, p(A) trial sequences), ensuring matched starting states and isolating learning-rule effects.

**Experimental design and procedure**

***Baseline generalization (Figure 3):*** With fixed latencies ([30, 90] ms), we evaluated cue-only predictions across 11 probability conditions (p(A) = 0–100% in 10% steps). We show the results for the final test trial for each condition.

***Single-block performance (Figure 4):*** We surveyed representative ISI configurations (e.g., [30, 90], [60, 60], [90, 30] ms) crossed with p(A) ∈ {50%, 80%, 100%} to test stationary learning across timing–probability pairs.

***Two-block adaptation (Figure 5):*** After 100 trials, parameters switched abruptly. We tested timing reversals ([30, 90]→[90, 30] ms) under p(A)=50% and 80%, as well as probability flips (e.g., 100%→0%, 80%→20%, 60%→40%) under [30, 90] ms.

***Multi-block adaptation (Figure 6):*** Four consecutive 100-trial blocks (switches at trials 100, 200, 300) alternated probabilities and ISIs to probe durability and interference:

- Schedule 1:
  p(A)=80%→20%→80%→20% and ISI = [30,90]→ [30,90] → [90,30] → [90,30] ms
- Schedule 2:
  p(A)=80%→80%→20%→20% and ISI = [30,90]→ [90,30] → [30,90] → [90,30] ms
- Schedule 3:
  p(A)=100% and ISI = [30,120]→ [60,90] → [90,60] → [120,30] ms.

All blocks had 100 trials; ITI jitter (0–200 ms) reduced overfitting to precise sequences. Reservoir state was never reset between trials.

***Comparative performance (Figure 7):*** All four learning variants were evaluated on the same four-block schedule (alternating p(A) = 80%↔20% with ISI evolving ([30, 120]→[60, 90] →[90, 60] →[120, 30] ms), carrying weights forward across blocks with fixed hyperparameters after initial selection.

**Analysis metrics**

***RMSE (Figures 3-8):*** Performance was quantified by the root mean squared error (RMSE) between the fused prediction $\mathrm{z}_{\text{prediction object}}(t)$ and the ground-truth profile (block probability p(A) × ground-truth timing $\mathrm{y}_{\text{when}}(t)$, normalized to [0,1]). RMSE was computed from cue offset (0 ms) to +150 ms post-cue and averaged across both channels.

***Probability-scaling analysis (Figure 3B):*** For each of the 11 probability conditions (p(A) = 0%–100% in 10% steps), we computed the window-averaged predicted amplitude in the *a priori* target windows after cue offset for the very last trial during run (Channel A: 30–60 ms; Channel B: 90–120 ms). We then quantified how predictions scaled with probability using (i) Pearson correlation, (ii) ordinary least squares (OLS) regression of windowed amplitude on p(A) with two-sided tests on the slope, and (iii) a non-parametric monotonic trend test (Spearman rank) against p(A).

***Block-wise baseline definition:*** For all block analyses (Figures 4–7), we defined “pre-adaptation” session baseline as the mean RMSE over the first 10 trials of that session. These earliest trials capture the model’s initial, minimally trained state at beginning of the simulation session. Improvement was assessed by comparing the end-of-block RMSE (trial 100) to this baseline with one-sided tests (end-of-block < baseline).

***Temporal window for RMSE (relative to cue offset):*** We define cue offset as time zero and compute RMSE over a fixed 150-ms window spanning post cue offset. For the ISIs used here (30, 60, 90, 120 ms), the corresponding target intervals (30–60, 60–90, 90–120, and 120–150 ms after cue offset) fall entirely within this window. If extended post-target assessment is desired, the upper bound can be increased without changing the analysis.

***PCA of readout geometry (Figure 8):*** To assess whether learning factorizes the prediction object into separable "what" and "when" components, we analyzed readout weights. For each trial, we concatenated the weight vectors for identity and timing across both channels ($\Phi_{\text{what}}$, $\Phi_{\text{when}}$; 4 vectors × 300 neurons = 1,200 dimensions). We normalized each weight type separately to unit variance before concatenation to ensure that both components contribute equally.

PCA was then fit on single-block data to define a reference latent space, into which multi-block trajectories were projected. Stability was quantified as the Euclidean distance from each projected point to its condition-matched single-block endpoint (the last-trial weight vector for that condition), aggregated across 60 condition–block pairs (20 conditions × 3 blocks). We then compared RMSE between methods at selected trials within each block (trials 1, 10, and 100) across all multi-block runs, using one-sided two-sample tests (Proposed Model RMSE < Gated FORCE RMSE) to relate representational geometry to performance.

This analysis is conceptually aligned with demixed PCA (dPCA) approaches, which uncover task-variable-aligned axes in shared neural populations (Kobak et al., 2016). While we did not fit dPCA explicitly, applying unsupervised PCA in normalized weight space provides a stringent test for emergent factorization: if "what" and "when" are learned as near-orthogonal components, their trajectories should diverge along distinct principal axes.

**Simulation apparatus**

All simulations were conducted using Python 3.10+ with PyTorch 2.1+ framework on CUDA-enabled GPU systems for accelerated computation.

# Tables

### Table 1: Core design features of the proposed model

| Design feature | What it does in the model | Mechanics (implementation detail) | Why this choice (biological / computational rationale) |
|---|---|---|---|
| **A) Heterogeneous Izhikevich reservoir** | Provides a stable, high-dimensional dynamical substrate with diverse intrinsic timescales so a brief cue unfolds into linearly decodable trajectories. | 1,000 Izhikevich neurons (≈80% E / 20% I); sparse recurrent matrix Ω with Dale-consistent signs; double-exponential synapses; 1 ms integration; small tonic bias $I_{\text{bias}}(t)$. Neuron parameters (a,b,c,d) are heterogeneously sampled to include regular-spiking, fast-spiking, and bursting units. | Mirrors cortical microcircuit diversity and supports echo-/liquid-state computation: mixed selectivity + varied time constants furnish a temporal basis on 10–100 ms scales; sparse, moderately strong recurrence keeps the system at an “edge-of-rich-dynamics” regime that is expressive yet controllable. |
| **B) “What” vs “when” separation in a shared readout population (multiplexing, not multi-stream)** | Two decoders read the same subset of reservoir neurons: $z_{\text{when}}(t)$ from instantaneous state $\boldsymbol{r}(t)$; $z_{\text{what}}$ from a post-cue average $\bar{\boldsymbol{r}}$. Functional separation with anatomical overlap. | Same binary mask $\phi_{\text{mask}}$ selects readout neurons; distinct weight matrices $\Phi_{\text{when}}$ and $\Phi_{\text{what}}$. | Treats identity and timing as separable functions over a shared mixed-selective code: the same neurons can carry combinations of variables, while weights carve out near-orthogonal readout axes. This is multiplexing (one population, multiple readouts), not multi-stream (separate modules), aligning with evidence for mixed selectivity and factorized manifolds in cortex. It boosts sample efficiency (shared features) yet avoids interference (separate readout weights). |
| **C) Online learning “when”** | Learns *during the trial* to align predictions with true latency windows. | **Online, gated 3-factor local Hebbian.** Each ms, update timing weights with the outer product of $\boldsymbol{r}(t)$ and the gated timing error $[y_{\text{when}}(t) - z_{\text{when}}(t)]$, scaled by the learning rate and masked by $\phi_{\text{mask}}$. | **Why online?** Timing is a temporal credit-assignment problem: the relevant teaching signal is localized in time (a narrow latency window). Updating within the trial lets eligibility traces overlap with the actual error, yielding fast, precise alignment of the predicted hazard around the true window and preventing drift. |
| **D) Offline learning “what”** | Learns *after* the trial from a stable post-cue state to consolidate identity. | **Offline, ungated 2-factor local Hebbian.** At trial end, update identity weights by the outer product of the post-cue averaged state $\bar{r}$ with the identity error $(y - z_{\text{what}})$; apply the same readout mask. | **Why offline?** Identity is trial-constant (the correct label doesn’t depend on within-trial time), so a post-cue average $\bar{r}$ is a low-variance sufficient statistic for the label. Deferring updates to trial end denoises momentary fluctuations and avoids the temporal credit problem entirely. |
| **E) Attention gate $\boldsymbol{G(t)}$ for “when”** | **Pre-detection:** permissive (learn calibrated “predict-zero” baselines for all channels).<br><br>**Post-detection:** selective (assign credit **only** to the | Binary/one-hot gate derived from teacher timing series and detected channel; applied only in the when rule; what uses G=1 offline. | **Why gated for ‘when’?** Timing errors are local in time and channel. Gating restricts plasticity to the detected channel/window, preventing cross-talk and enabling rapid, drift-free recalibration after switches, consistent with phasic neuromodulatory control and hitting the stability–plasticity sweet spot. |
| | detected channel/time window). | | **Why ungated for 'what'?** Because the target is constant across the trial, a gate adds no information and may inject unnecessary stochasticity; a simple 2-factor pre × error rule is enough and more biologically economical. This slow(er) consolidation complements fast timing updates, reducing interference and improving calibrated probability estimates. |
| **F) Timing-based feedback $I_{\text{feedback}}$** | Uses the model's own timing prediction to stabilize and phase-align reservoir dynamics without teacher forcing. | Inject $I_{\text{feedback}}(t) = E^{\top} z_{\text{when}}(t)$ with Dale-consistent signs and modest gain at each time step | Implements top-down gain control / recurrent stabilization, keeping trajectories near the correct latency manifold and preventing drift. |
| **G) Fusion of "what" × "when" (single prediction object)** | Combines identity and timing into one spatiotemporal prediction trace whose amplitude reflects probability and whose temporal profile aligns to the expected latency window. | Elementwise multiplicative fusion per channel: $\boldsymbol{z}_{fused}(t) = \boldsymbol{z}_{\text{what}} \odot z_{\text{when}}(t)$. "What" is trial-constant; "when" is time-resolved | A simple **multiplicative (AND-like)** fusion turns two decoders into a **single, complete prediction object**: identity sets *which* channel is expected (and with what confidence), while timing selects *when* that expectation should express. The product yields a **probability-weighted, time-resolved trace**, high only for the right channel at the *right moment,* so errors remain interpretable (identity vs. timing) and the two axes stay separable in weight space. This echoes **biological gain control**, where contextual belief multiplicatively gates time-locked anticipatory signals. |

**Table 2: Primary testable hypotheses**

| Hypothesis | Claim | Prediction | Test | Falsifier |
|---|---|---|---|---|
| **A) Prediction error as modulator, not feedforward driver** | Prediction errors (PEs) modulate local plasticity and feedback gain but do not drive feedforward activity. Prediction-driven signals form the effective feedforward flow through the network. | (1) PE magnitude predicts next-trial adaptation but not same-trial cue-driven prediction. (2) Removing or flattening PE gating slows adaptation but preserves cue-driven propagation. (3) Feedforward channels are multiplexed, carrying prediction and transient PE components. | **Human**: MEET with alternating trained/untrained blocks; regress Δprediction on preceding PE. Use spectral dissociation (beta/alpha = prediction; theta/gamma = PE) and autonomic proxies (pupil, skin conductance) for modulatory gain. **Rodent**: Neuropixels during temporal-expectation task; relate PE-locked spikes/LFPs to next-trial decoder-weight changes and confirm multiplexed feedforward signatures. | 1) PE magnitude correlates with same-trial cue-driven prediction amplitude. (2) No spectral or temporal separation between prediction and PE. (3) Removing PE gating abolishes predictive propagation, implying PE drives rather than modulates feedforward activity. |
| **B) Orthogonal "what" vs "when" subspaces** | Mixed-selective populations encode separable identity and timing axes within a shared neural manifold. Orthogonality allows flexible reuse of the same population under changing contexts. | Component or subspace analysis (e.g., dPCA/CCA) reveals near-orthogonal axes for identity ('what') and latency ('when')." After context switches, trajectories return rapidly to condition-specific endpoints without collapse. | **Human**: MEG/ECoG during MEET; subspace analyses on population activity or fitted decoders. **Rodent**: Neuropixels recordings; compare latent geometry and within-block RMSE to FORCE-like baseline fits to assess representational separability. | Identity and timing cannot be decoupled, or trajectories collapse centrally after switches, falsifying the claim that mixed-selective populations support multiplexed, orthogonal subspaces. |
| **C) Intrinsic dynamics shape prediction** | Predictions emerge from the interaction of cue-evoked and intrinsic dynamics; altering internal timescales changes predictive precision and temporal alignment. | Increased cortical variability (drowsiness, low-dose ketamine) broadens timing distributions; heightened arousal (caffeine) sharpens precision, particularly under weak priors or uncertain intervals. | **Human**: MEET with arousal manipulations; relate Fano factor and state-space dispersion to timing RMSE. **Rodent**: Modulate background drive pharmacologically; assess how reservoir-like state variance alters temporal precision and adaptation speed. | Prediction precision remains invariant across intrinsic-state changes, indicating that timing accuracy does not depend on intrinsic dynamics. |

**Table 3: Secondary testable hypotheses**

| Hypothesis | Claim | Prediction | Test | Falsifier |
|---|---|---|---|---|
| **A) Phase-specific neuromodulatory gating** | Brief, phase-locked ACh/DA/NE bursts gate "when" plasticity post-detection; "what" consolidates offline without the gate. | Enhancing ACh (donepezil) selectively speeds "when" adaptation within blocks; identity learning changes little within-trial but improves gradually across trials. | **Human**: MEET + EEG + pupillometry under placebo/donepezil/scopolamine. **Rodent**: Opto/pharmacological modulation of basal forebrain or VTA inputs during post-detection window. | If modulators accelerate "what" learning equally or timing improves without gating, phase-specificity fails. |
| **B) Closed-loop feedback stabilizes timing manifolds** | Top-down prediction feedback ($I_{\text{feedback}}$) phase-aligns dynamics, reducing latency jitter without teacher forcing. | Disrupting feedback increases latency jitter and weakens "return-to-manifold" after switches. | **Human**: MEET + EEG + pupillometry under placebo/donepezil/scopolamine. **Rodent**: Opto/pharmacological modulation of basal forebrain or VTA inputs during post-detection window. | If feedback disruption leaves timing precision unchanged, stabilization is not feedback-dependent. |
| **C) Multiplicative fusion (what × when) as cortical gain** | Final prediction arises from multiplicative gating of timing ramps by identity confidence. | Lower cue confidence scales down amplitude without shifting latency; higher p(A) boosts amplitude with minimal latency change. | **Human**: MEET with cue-ambiguity or prior-probability manipulations; compare amplitude vs latency effects in cue-only responses; look for gain-like scaling. | If ambiguity alters latency as much as amplitude, interaction is not multiplicative. |
| **D) Pre-detection "predict-zero" baseline learning** | Before channel detection, gating supports a calibrated "no-event-yet" baseline shared across channels. | CNV-like baseline emerges pre-target, becoming channel-specific only post-detection; ACh boost sharpens post-detection selectivity without inflating baseline. | **Human**: EEG CNV analysis under pharmacological modulation. **Rodent**: LFP/ensemble divergence only after teacher onset. | If channel selectivity exists before detection with equal priors, a two-phase gate is unnecessary. |

## Figures

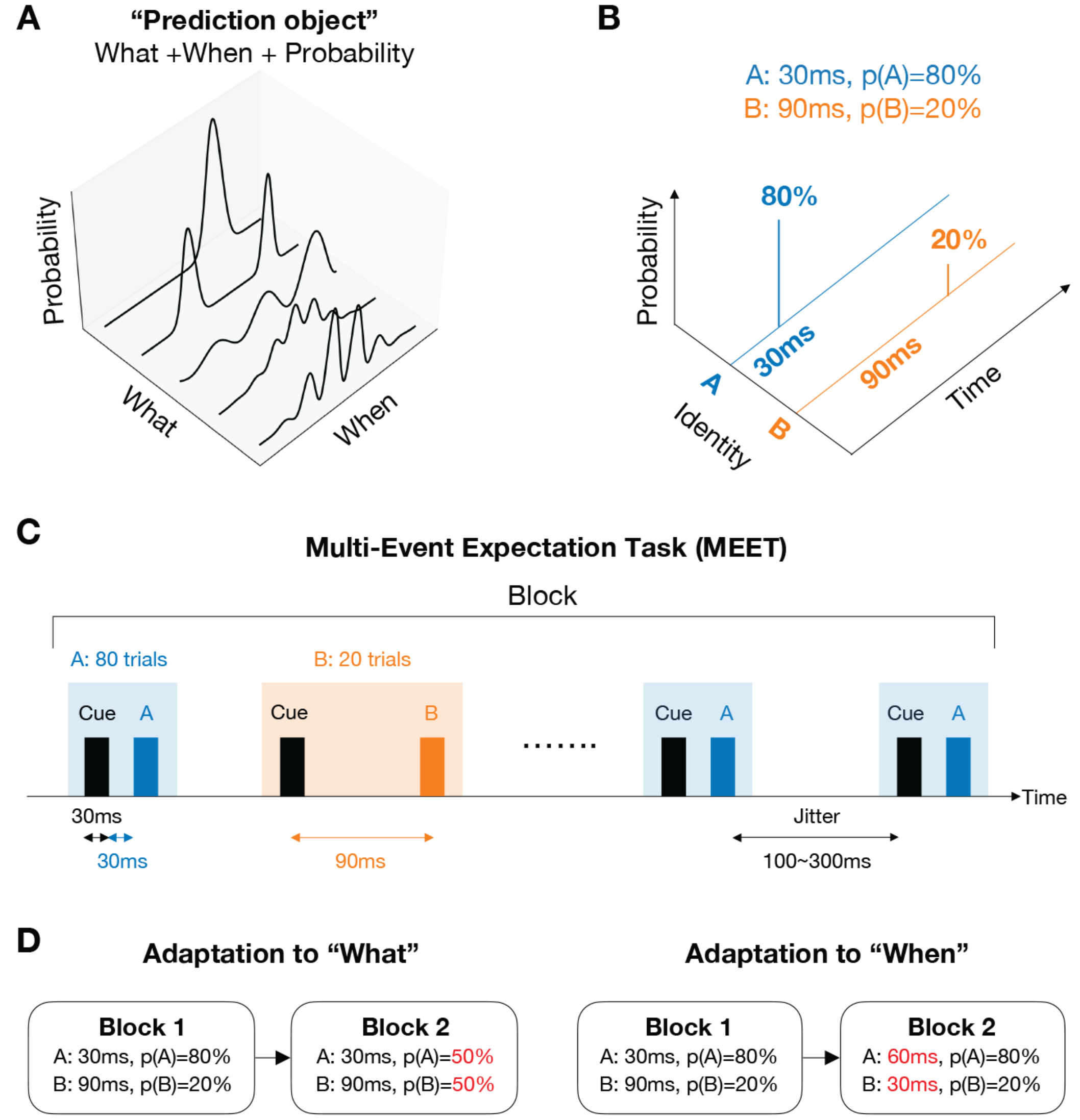

**Figure 1: Multi-Event Expectation Task (MEET)– design and experimental paradigm.**
**A,** Three-dimensional prediction space showing the dual-task requirements for "what" (channel identity: A vs B) and "when" (temporal prediction) with associated probability distributions. **B,** Task parameter space illustrating two example conditions: Channel A emerges at +30ms post-cue offset with probability p(A)=80% (green), while Channel B emerges at +90ms post-cue offset with probability p(B)=1-p(A)=20% (blue). The model must predict both which channel will respond ("what") and when that response will occur ("when"). **C,** Temporal structure of a single trial showing cue presentation, variable delay periods, and block organization. Trials are organized into blocks of 100 trials each with consistent ISI parameters. **D,** Adaptation paradigm demonstrating block-wise changes in both “what” (channel probabilities p(A), p(B)) and “when” (channel-specific temporal intervals) parameters across consecutive blocks, testing the model’s ability to track changing environmental statistics.

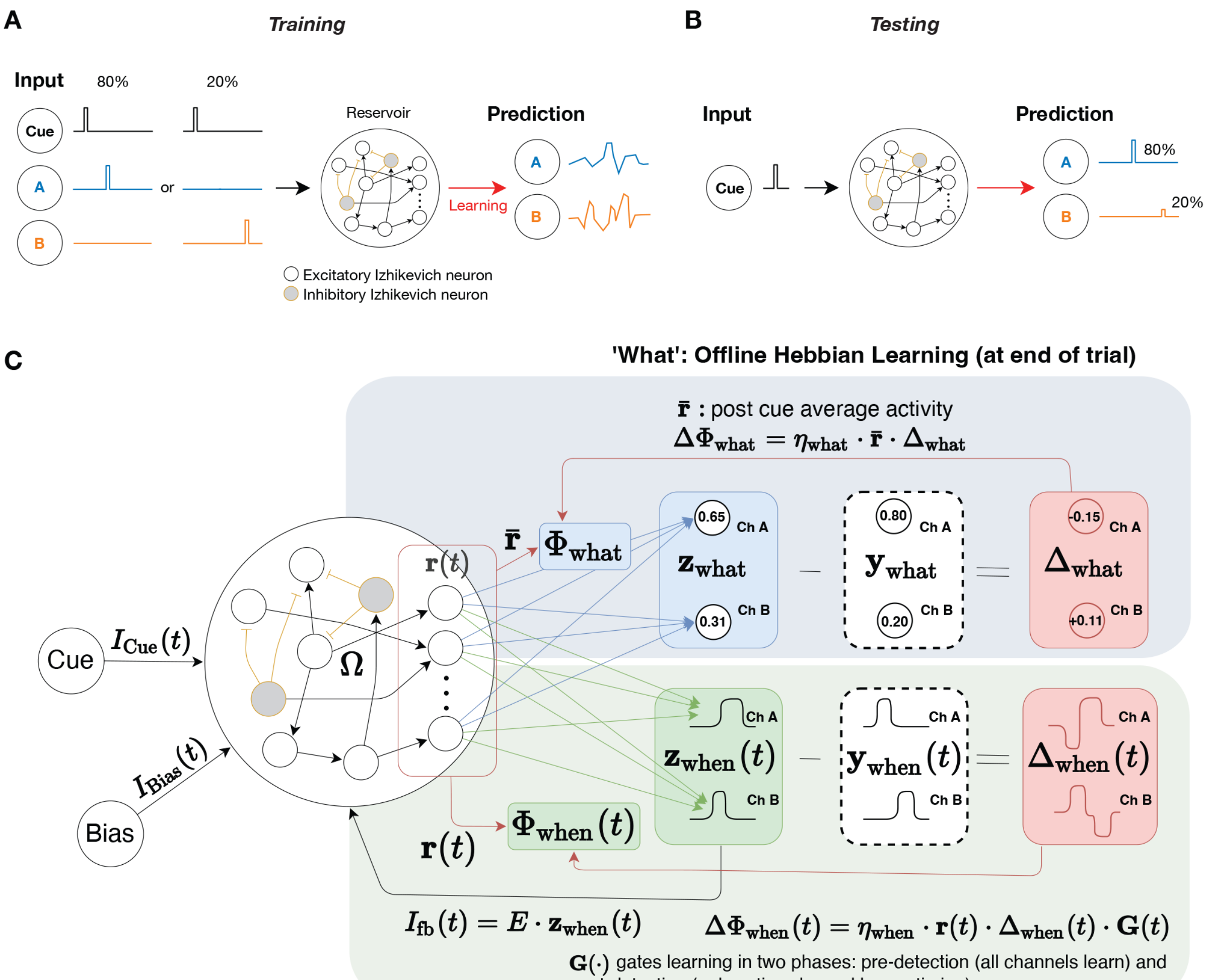

**Figure 2: Proposed model– gated local learning model architecture.**

**A,** Training phase. During training, the reservoir receives three inputs: a brief Cue and one of two mutually exclusive stimuli (Channel A or Channel B). The identity of the stimulus is sampled according to the block probability (e.g., 80% A, 20% B), and each stimulus occurs at its channel-specific latency. The recurrent spiking reservoir transforms these inputs into high-dimensional activity, and learning adjusts only the readout weights so that the network can predict both which channel will occur ("what") and when it will occur ("when"). **B,** Testing phase (cue-only prediction). After learning, the Cue alone is presented and no stimulus is delivered. If the model has successfully acquired the complete prediction object, the Cue evokes anticipatory activity in both channels with amplitudes proportional to their learned probabilities (e.g., 80% for A, 20% for B) and temporally aligned to their expected latencies. Thus, the network generates a probability-weighted, time-resolved prediction in advance of any external input. **C,** Circuit diagram of the biologically inspired spiking-reservoir model. A 1,000-neuron Izhikevich reservoir receives cue input $I_{\text{cue}}(t)$, bias $I_{\text{bias}}(t)$, recurrent drive $\Omega$, and sparsely pre-fixed, Dale's-law-consistent timing feedback $I_{\text{feedback}}(t) = E^{\top} z_{\text{when}}(t)$ from the current timing prediction. Two readouts share the same sparse pool of reservoir neurons: the identity ("what", blue) decoder operates offline from the post-cue average state and learns by an ungated two-factor Hebbian rule (presynaptic activity × identity error), whereas the timing ("when", yellow) decoder operates online at 1-ms resolution and learns by a gated three-factor Hebbian rule (presynaptic activity × timing error × gate). The gate $G(t)$ implements a two-phase schedule: pre-detection (permissive across channels to calibrate predict-zero baselines) and post-detection (plasticity restricted to the active channel within its latency window). During evaluation, only the cue is provided and learning is disabled; the timing prediction output is fed back to stabilize phase without leaking ground truth. This dual-pathway, locally learned architecture acquires the complete prediction object (what × when × probability) without backpropagation or global error broadcast.

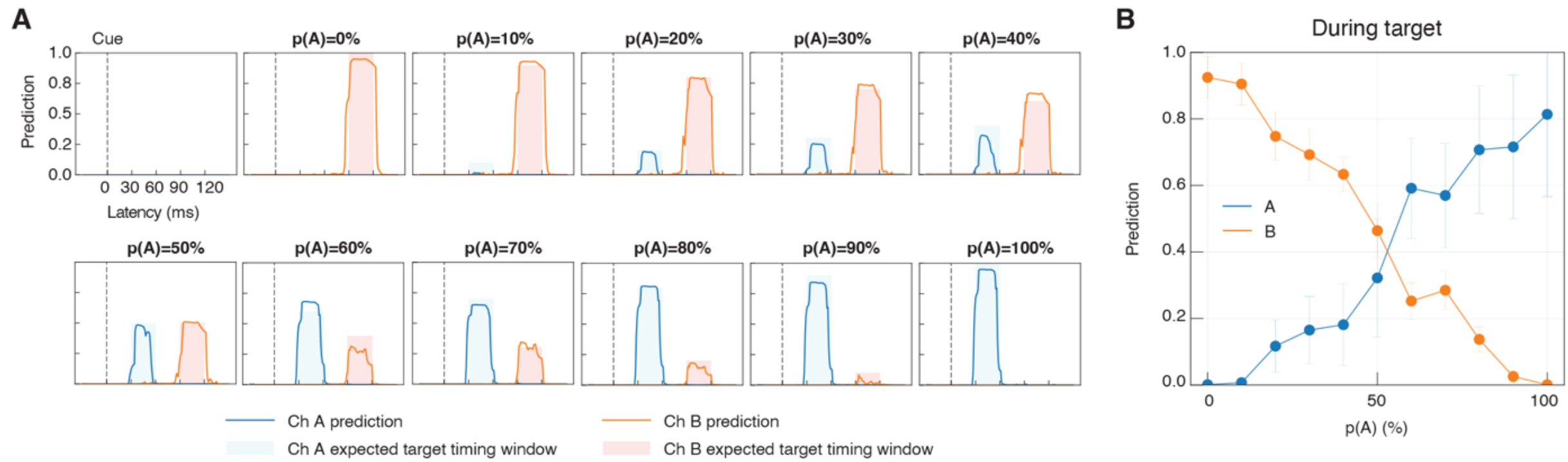

**Figure 3: Robust event prediction and probability tuning.**

**A,** Test performance evaluation across different channel probabilities (p(A) values from 0% to 100%) for fixed channel timing conditions (Channel A at 30 ms post-cue offset, Channel B at 90 ms post-cue offset, with 30 ms stimulus duration for both channels). Each subplot shows the combined 'what' × 'when' predictions as temporal prediction traces from the final test trial in which only the cue was presented, with Channel A predictions (blue) and Channel B predictions (orange) plotted over time relative to cue offset. Shaded regions indicate the expected target timing windows for each channel and the height corresponding to p(A) (30-60 ms for Channel A in light blue, 90-120ms for Channel B in light orange). **B,** Average predicted values during the expected target timing windows across all p(A) conditions. Blue line shows mean Channel A predictions during its 30-60ms target window, orange line shows mean Channel B predictions during its 90-120 ms target window, with error bars representing standard deviations.

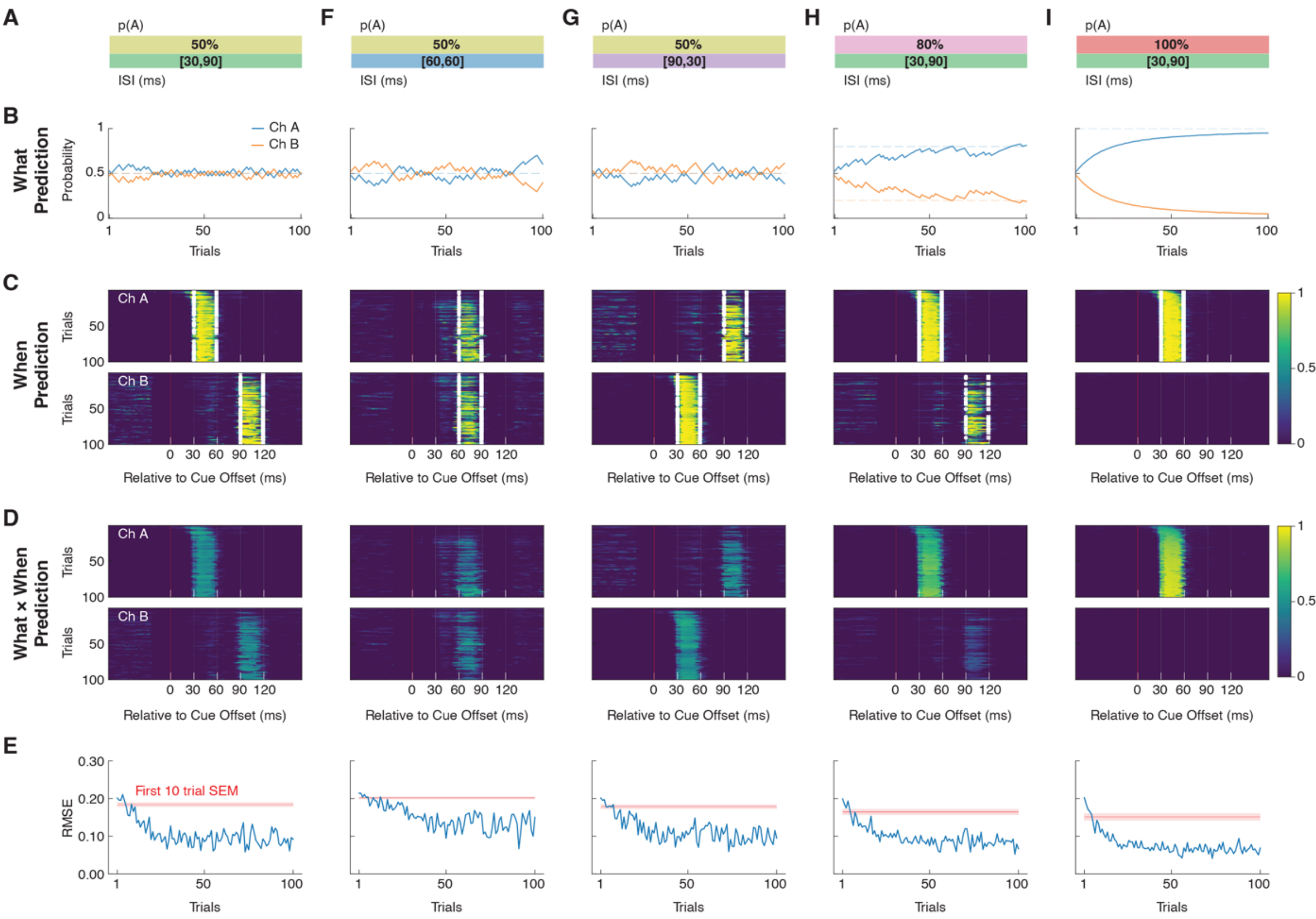

**Figure 4: Single-block stationary performance.**

Comprehensive evaluation of model performance across single-block conditions with specified channel probability p(A) and interstimulus interval (ISI). **A,** Block design for the representative base condition used in panels B–E: p(A)= 50% with ISI=[30,90] ms over 100 trials. **B,** Channel identity ("what") predictions for the base condition, showing trial-by-trial estimates for Channel A (blue) and Channel B (orange). **C,** Temporal ("when") predictions for the base condition as a heatmap (trials on y-axis; time relative to cue offset on x-axis); the red vertical line marks cue offset and white markers indicate ground-truth onset/offset per trial. **D,** Combined what × when predictions for the base condition, in the same heatmap format as C. **E,** RMSE for the base condition across trials; the solid red horizontal line is the mean of the first 10 trials (±1 SEM shaded). **F–I**, Additional single-block conditions (each following the same analysis as **B–E**) shown explicitly: **F,** p(A)= 50%, ISI=[60,60] ms; **G,** p(A)=50%, ISI=[90,30] ms; **H,** p(A)=80%, ISI=[30,90] ms; **I,** p(A)=100%, ISI=[30,90] ms.

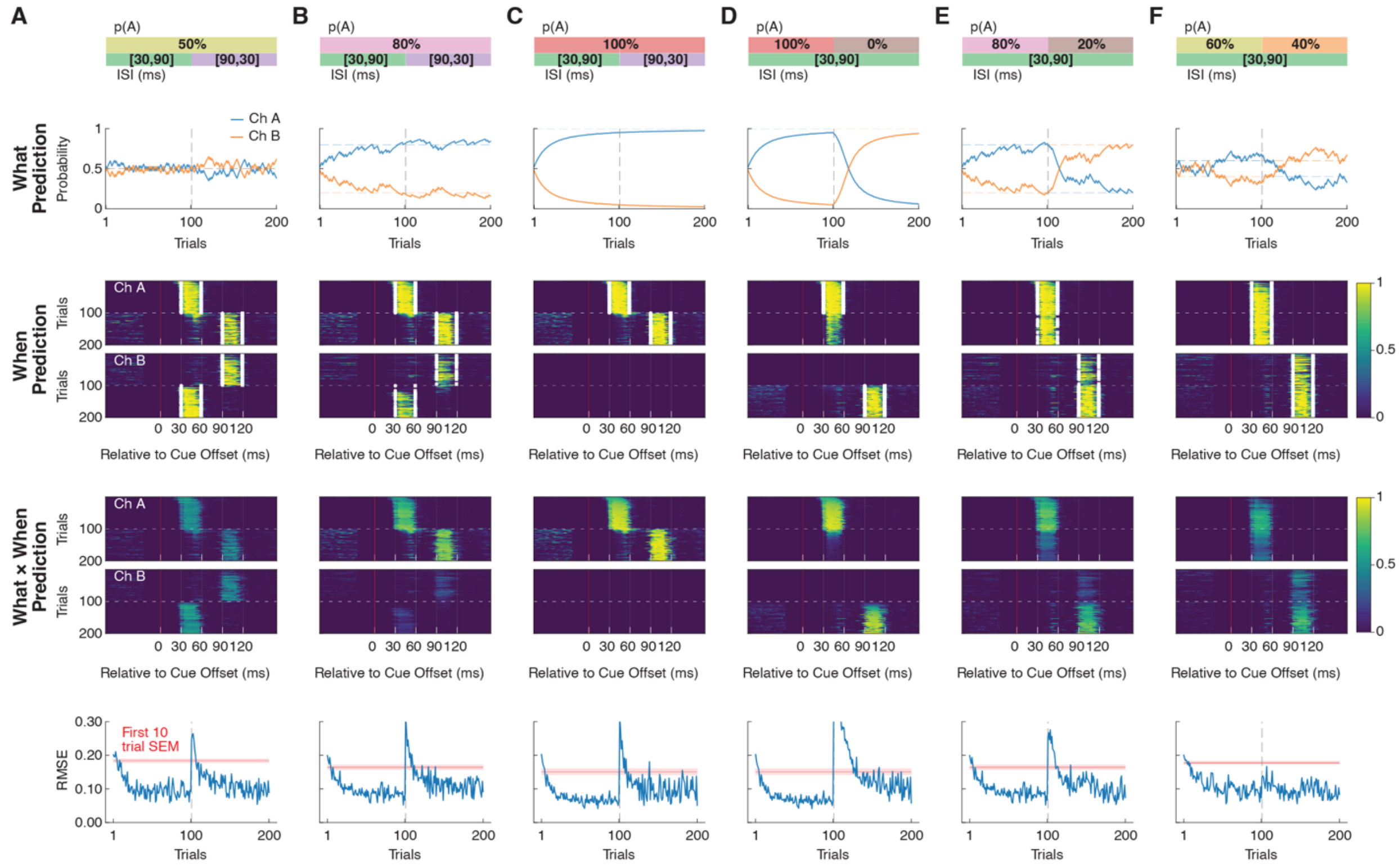

**Figure 5: Adaptation to changing task conditions.**

Model performance during block transitions demonstrating rapid adaptation to environmental changes in both temporal and probability domains. **A,** Two-block sequences with transitions after trial 100 (ISI: [30,90]→[90,30] ms at p(A)=50%). **B,** ISI: [30,90]→[90,30] ms at p(A)=80%, **C:** ISI: [30,90]→[90,30] ms at p(A)=100%, and probability inversions under ISI = [30,90] ms (**D:** p(A): 100%→0%, **E:** 80%→20%, **F:** 60%→40%). Matching colors indicate identical p(A) and ISI parameter combinations across blocks. Conventions used in each subplot (what prediction, when prediction, fused what × when prediction, and RMSE) follow Figure 4.

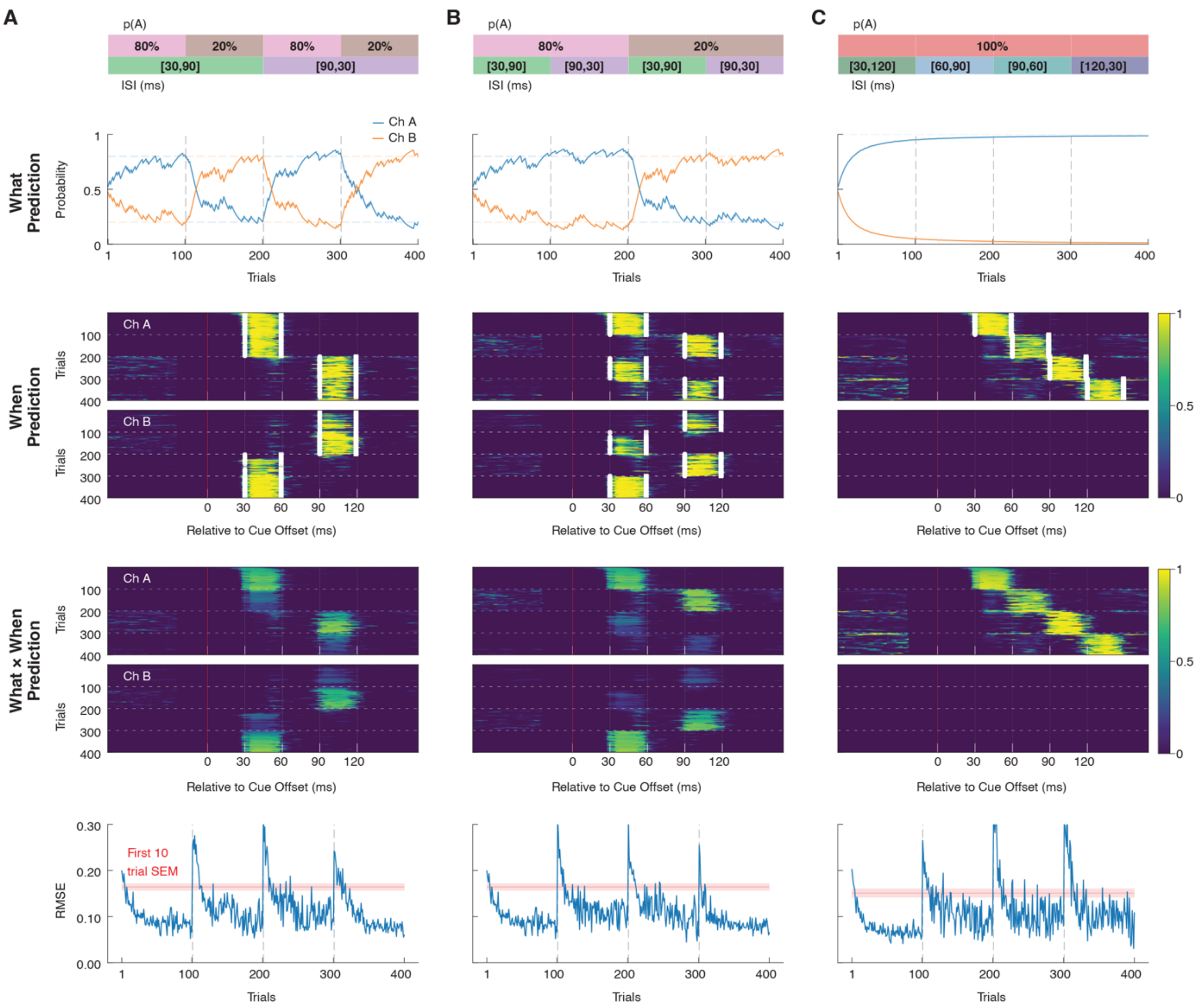

**Figure 6: Multi-block adaptation across extended sequences.**

Evaluation of sustained adaptive performance through multiple consecutive environmental changes, testing the model's capacity to track evolving task statistics. **A,** Schedule 1 condition: Unified block design featuring a four-block sequence with alternating probability (p(A): 80%→ 20%→80%→20%) and timing reversal (ISI: [30,90]→[30,90]→[90,30]→[90,30] ms). **B,** Results for Schedule 2 condition: Unified block design featuring a four-block sequence with alternating probability (p(A): 80%→80%→20%→20%) and timing reversal (ISI: [30,90]→[90,30]→[30,90]→ [90,30] ms). **C,** Results for Schedule 3 condition: Unified block design featuring a four-block sequence with fixed probability (p(A): 100%) and timing reversal (ISI: [30,120]→[60,90]→[90,60] →[120,30] ms). Conventions used in each subplot (what prediction, when prediction, fused what × when prediction, and RMSE) follow Figure 4. White dashed horizontal lines indicate block transitions.

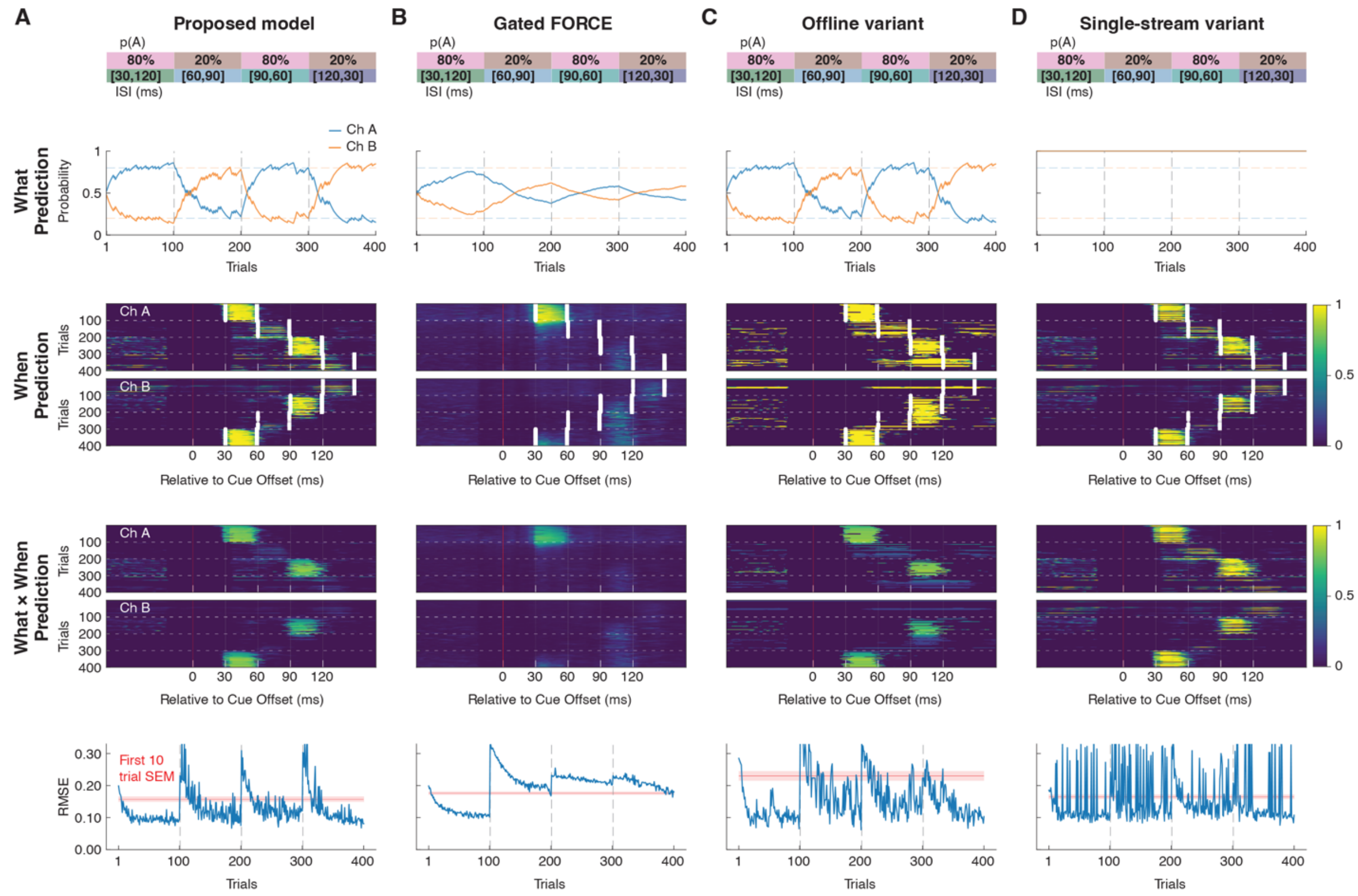

**Figure 7: Comparative performance across learning algorithms.**

Systematic comparison of different learning paradigms under identical multi-block adaptation challenges, revealing fundamental differences in adaptive capacity and stability. **A,** Results for the Proposed model, featuring a four-block sequence with alternating probability (p(A): 80%→ 20%→80%→20%) and timing reversal (ISI: [30,120]→[60,90]→[90,60]→[120,30] ms). **B,** Results for Gated FORCE model. **C,** Results for Offline variant model (both 'what' and 'when' are offline) learning. **D,** Results for Single-stream variant model (no separation between what and when) learning. Conventions used in each subplot (what prediction, when prediction, fused what × when prediction, and RMSE) follow Figure 4. White dashed horizontal lines indicate block transitions.

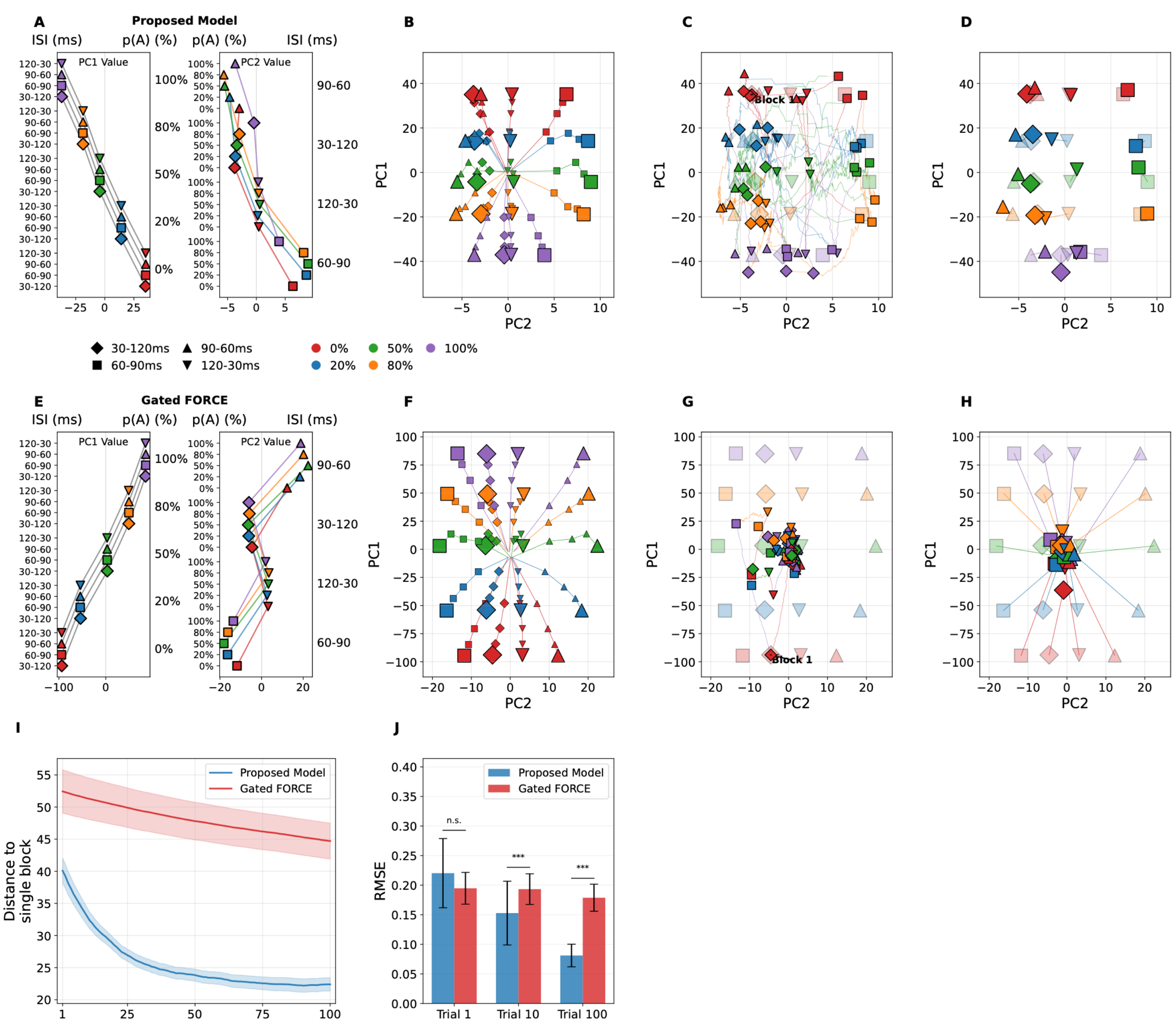

**Figure 8: Principal component analysis of readout weight dynamics.**

**A,** Principal components for the Proposed model from single-block analyses. The left subplot shows PC1 values across four ISI configurations grouped by channel probability p(A) (20 conditions total); gray lines connect identical ISI configurations across probabilities. The right subplot shows PC2 values across five p(A) levels grouped by ISI configuration; colored lines connect identical p(A) levels. ISI order determined based on group average for visualization purposes. Shape and color encodings follow the legend. **B,** Trial-by-trial trajectories in the PC1–PC2 plane for single-block runs of the Proposed model. Points mark within-block progression (25%, 50%, 75%) with larger markers at the final (100%) trial of each block. **C,** Multi-block adaptation trajectories in the PC1–PC2 plane for the Proposed model run with 20 pseudo-randomly ordered conditions repeated three times. Transparent markers indicate the corresponding single-block reference positions as shown in B; Solid markers highlight the final (100th) trial of each block. **D,** Average latent position (PC1–PC2) for each condition in the multi-block run, averaged across the three condition-matching repeats. Faint (transparent) markers show the corresponding single-block endpoints from B. For each condition, a line links the multi-block mean to its single-block reference, visualizing any offset in latent space. **E,** Principal components for the Gated FORCE model from single-block analyses, formatted as in panel A. ISI group ordering used in the right subplot follow that of A. **F,** Trial-by-trial trajectories in the PC1–PC2 plane for single-block runs of the Gated FORCE model, formatted as in panel B. **G,** Multi-block adaptation trajectories in the PC1–PC2 plane for a single Gated FORCE run with the same condition ordering as panel C; transparent markers indicate single-block reference positions and markers indicate final trials per block. **H,** Average latent positions (PC1–PC2) for Gated FORCE multi-block conditions, shown with transparent single-block references and connecting lines as in panel D. **I,** Mean Euclidean distance (± dispersion) between each multi-block point and its single-block reference, computed across 60 condition–block pairs (20 conditions × 3 repeated blocks, with condition order in each repeat pseudo-randomized from a fixed seed) and plotted as a function of within-block trial index; blue denotes the Proposed model and red denotes Gated FORCE. **J,** Root-mean-square error (RMSE) at trials 1, 10, and 100 within blocks during multi-block adaptation. Bars show mean RMSE across 60 blocks with error bars indicating standard deviation; bracket annotations indicate significance labels (*** $p < 0.001$, ** $p < 0.01$, * $p < 0.05$, n.s.).

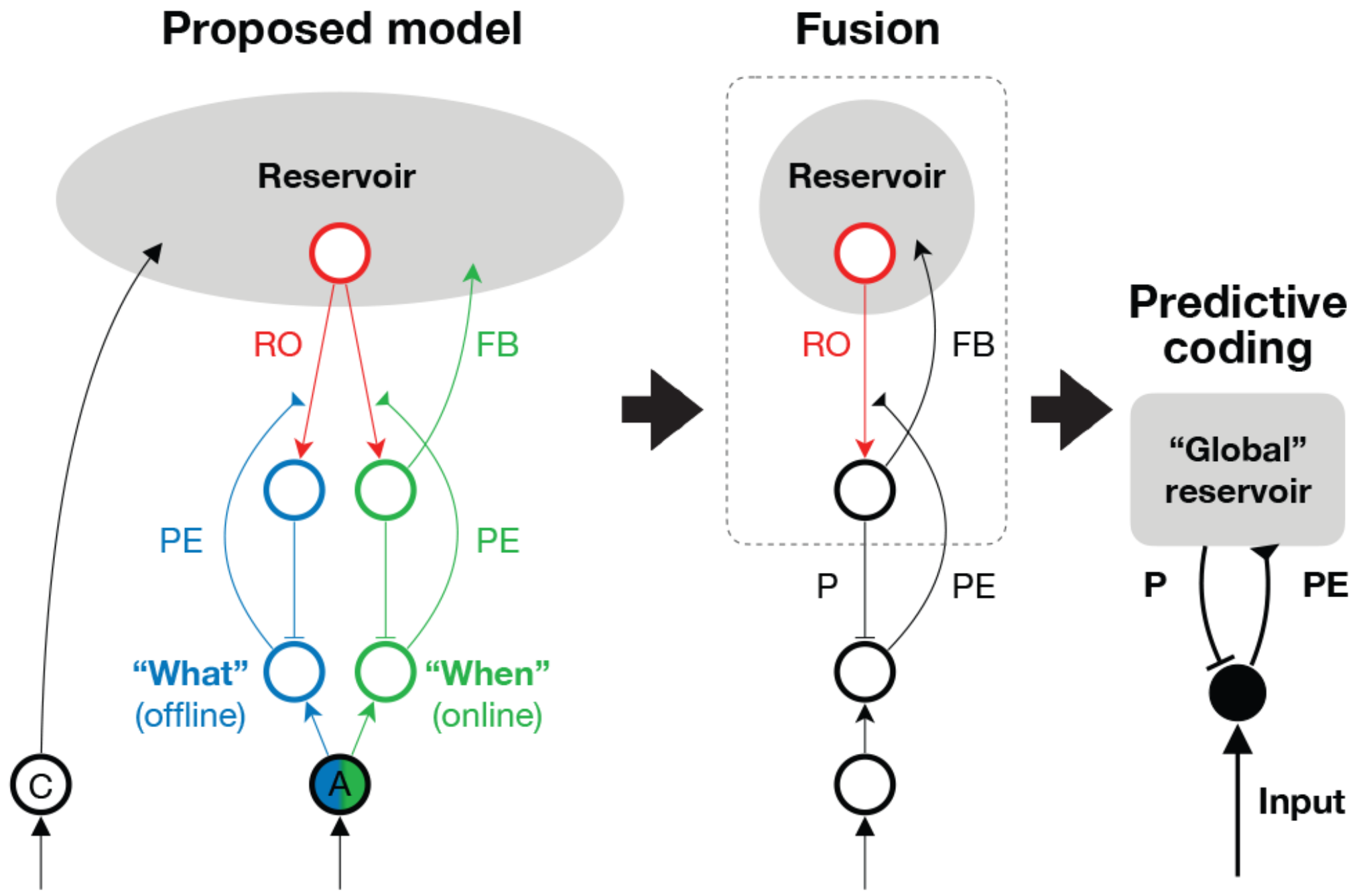

**Figure 9: Proposed model as a predictive-coding framework.**

Conceptual diagram illustrating the proposed model (left), with the B stream omitted for simplicity. The "what" and "when" error computations are shown in blue and green, respectively, and the readout is indicated in red. RO: readout; PE: prediction error; FB: feedback. A simplified single-stream version (middle) shows fused "what" and "when" components. When the module within the dashed box is interpreted as a "global" reservoir, the model can be viewed as a predictive-coding framework (right).

**Data availability**

The data supporting the findings of this study are available from the corresponding author upon request. Raw simulation outputs, processed data, and analysis scripts used to generate all figures will be made fully available to editors and reviewers upon submission and will be publicly released upon publication.

**Code availability**

All code used in this study are made fully available to editors and reviewers upon submission and will be publicly released upon publication. The computational framework is implemented in Python 3.10+ with PyTorch 2.1+. Detailed installation instructions and scripts to reproduce all figures are provided in the repository.

**Ethics statement**

This study involved only computational modeling and simulation. No human participants or animal subjects were involved.

**Reporting Summary**

Further information on research design is available in the link.

**Acknowledgements**

We thank Dr. Eugene Izhikevich, Dr. Pawel Herman and Dr. Heng Zhang for helpful discussions and comments. Computational resources were provided by IRCN. This work was supported by World Premier International Research Center Initiative (WPI), MEXT, Japan (to Z.C.C.).

**Author contributions**

Z.C.C. conceptualized the research idea and provided overall guidance. Y.Y. wrote the code, implemented the computational methods, and conducted the simulations. Y.Y. and Z.C.C. co-wrote the manuscript. Both authors contributed to manuscript revision and approved the final version.

**Competing interests**

The authors declare no competing interests.

**Correspondence**

Correspondence should be addressed to Z.C.C. and Y.Y.

## Supplementary Materials

### Supplementary Method 1: Systematic assessment of forgetting in recursive least squares learning

To test whether limited adaptability in Gated FORCE could be attributed to insufficient forgetting of past statistics, we introduced a forgetting factor into the recursive least squares update and systematically varied it systematically across a broad range. This manipulation tests whether excessive retention of earlier correlations constrains rapid reconfiguration when task statistics change.

In standard recursion least squares with $\lambda_f$ = 1.0, the inverse correlation matrix accumulates information across all past trials and progressively stabilizes the solution. This property is advantageous under stationary conditions but can hinder rapid adaptation because earlier correlations continue to influence subsequent updates. Introducing a forgetting factor $\lambda_f$ < 1 discounts older information and increases responsiveness, allowing us to test whether reduced adaptability reflects excessive memory of past statistics.

We evaluated five variants spanning minimal to moderate forgetting, with $\lambda_f \in$ {1.0, 0.999999, 0.99999, 0.9999, 0.999}, alongside the proposed model (Supplementary Figure 1A). Values extremely close to 1 approximate standard recursive least squares with very slow decay of past statistics, whereas smaller values increase plasticity at the cost of greater variance, reflecting the classical stability-plasticity trade-off in adaptive filtering. Moderate forgetting produced asymptotic errors similar to standard recursive least squares, whereas stronger forgetting increased variability and degraded overall performance. Across all settings, no value of $\lambda_f$ restored the rapid convergence or stable synaptic geometry observed in the proposed model.

Forgetting increased short-term responsiveness but did not prevent geometric collapse in weight space. Even when smaller $\lambda_f$ values accelerated early error reduction, trajectories in principal component space drifted away from condition-specific single-block reference configurations rather than reconverging onto them. Because the single-block and multi-block simulations were conducted independently with identical initial seeds, this comparison reflects convergence toward characteristic solutions rather than reinstatement of stored weights. The failure to reconverge therefore indicates that the limitation of Gated FORCE is not simply excessive retention of past statistics.

Instead, the results point to a structural property of globally coupled least-squares updates. Each error update reshapes the entire readout subspace, so learning for one condition perturbs representations associated with others. This global coupling leads to entanglement of condition-specific synaptic configurations and prevents stable re-formation of separable axes after switches.

In contrast, the proposed model employs local, attention-gated plasticity that restricts updates to synapses associated with the active channel and latency window. This structured credit assignment preserves separable what and when dimensions while still allowing rapid reassignment under changing statistics. Together, Supplementary Figures 1 show that robust multidimensional prediction depends on organized synaptic subspaces rather than simply increasing the rate of forgetting.

## Supplementary Method 2: P Matrix Initialization Sensitivity Analysis for Gated FORCE learning

### Overview

To assess the robustness of the Gated FORCE learning rule to hyperparameter choices, we systematically varied the $P$ matrix initialization parameters $\alpha_{when}$, $\alpha_{what}$ that control the initial $P$ matrix scaling for the temporal $z_{when}$ and identity $z_{what}$ readouts, respectively.

### P Matrix in FORCE Learning

FORCE learning uses the recursive least squares (RLS) algorithm to update readout weights. Unlike Hebbian learning rules that use an explicit learning rate parameter ($\eta$) to scale weight updates, FORCE learning has no such parameter. Instead, the $P$ matrix that approximates the inverse of the input correlation matrix entirely determines the magnitude of weight updates. The weight update rule is:

$$\Delta w = -e * P * r$$

where $e$ is the prediction error, $r$ is the network activity vector, and $P$ modulates how strongly the error drives weight changes.

The P matrix is initialized as:

$$P_0 = I/\alpha$$

where $I$ is the identity matrix and $\alpha$ is a scalar initialization parameter. Because FORCE has no explicit learning rate, $\alpha$ is the primary parameter controlling learning dynamics through its effect on the initial $P$ matrix:

1. Smaller $\alpha$ values produce larger initial $P$ matrices ($P_0 = I/\alpha$), resulting in larger weight updates and faster initial learning. However, this can lead to instability or overshooting.
2. Larger $\alpha$ values produce smaller initial $P$ matrices, resulting in smaller weight updates and slower, more conservative learning. This provides stability but may require more trials to converge.

During learning, $P$ is updated according to the RLS rule with forgetting factor $\lambda_f$:

$$P_{new} = \left(\frac{1}{\lambda_f}\right) * \left[P - \left(\frac{P * r * r' * P}{\lambda_f + r' * P * r}\right)\right]$$

When no forgetting is applied ($\lambda_f = 1$), this reduces to the standard RLS update:

$$P_{new} = P - \left(\frac{P * r * r' * P}{1 + r' * P * r}\right)$$

Without forgetting, $P$ gradually decreases as the network accumulates information about input statistics, naturally reducing the magnitude of weight updates over time. With forgetting ($\lambda_f < 1$), the $1/\lambda_f$ scaling factor causes $P$ to grow, preventing this decay and allowing the network to continue adapting to changing statistics.

In our model, separate $P$ matrices are maintained for the temporal readout ($z_{when}$, updated at each timestep) and identity readout ($z_{what}$, updated once per trial), with initial scaling controlled by $\alpha_{when}$ and $\alpha_{what}$ respectively.

**Gated Learning for the P Matrix**

The Gated FORCE model implements selective learning through a signal-detection gating mechanism that affects both weight updates and $P$ matrix updates. The gating operates as follows:

1. Before any teacher signal is detected in a trial, the learning gate is open for all output channels, allowing the network to learn to predict baseline activity (zero output).
2. Once a teacher signal is detected, the learning gate restricts updates to only the active output channel (determined by which cue was presented).

This gating is applied by scaling the prediction error $e$ before computing the $P$ matrix update. Since the $P$ matrix update depends on the error-weighted gain $k = P * r / (\lambda_f + r' * P * r)$, gating the error effectively gates the entire RLS update including both the weight change and the $P$ matrix modification. This ensures that the $P$ matrix only accumulates input statistics from trials where learning is relevant for each output channel, preventing interference between the two readout pathways.

**Experimental Conditions**

We tested five combinations of initialization parameters:

1. ($\alpha_{when} = 1, \alpha_{what} = 10$): Fast temporal learning, normal identity learning
2. ($\alpha_{when} = 10, \alpha_{what} = 10$): Default configuration (baseline)
3. ($\alpha_{when} = 100, \alpha_{what} = 10$): Slow temporal learning, normal identity learning
4. ($\alpha_{when} = 10, \alpha_{what} = 1$): Normal temporal learning, fast identity learning
5. ($\alpha_{when} = 10, \alpha_{what} = 100$): Normal temporal learning, slow identity learning

**Simulation Protocol**

- **Single-block simulations.** For each $\alpha$ condition, we ran 20 independent simulations covering all combinations of 5 probability conditions (p(A) in {0, 0.2, 0.5, 0.8, 1.0}) and 4 inter-stimulus interval (ISI) patterns ({30-120, 60-90, 90-60, 120-30} ms). Each simulation consisted of 100 trials within a single block. Network weights were randomly initialized with a fixed seed to ensure reproducibility across conditions.
- **Multi-block simulations.** To evaluate adaptation performance, we ran 60-block simulations for each $\alpha$ condition using randomized block sequences. Each block contained 100 trials with a unique combination of p(A) and ISI values, presented in a pseudo-random order (seed=42) to ensure identical block sequences across conditions.

**Baseline Comparison**

The Gated Hebbian model (proposed model) was included as a reference to contextualize the Gated FORCE variants' performance. The default Gated FORCE configuration ($\alpha_{when} = 10$, $\alpha_{what} = 10$) corresponds to the same model used in the main analyses (Figure 8).

**Implementation Details**

All simulations used identical network architecture (N=1000 Izhikevich neurons), random weight initialization, and deterministic data streaming to ensure reproducibility. Simulations were performed on CUDA-enabled GPUs with checkpoint-based memory management for the multi-block runs.

## Supplementary Figure 1: Sensitivity of Gated FORCE to forgetting factors

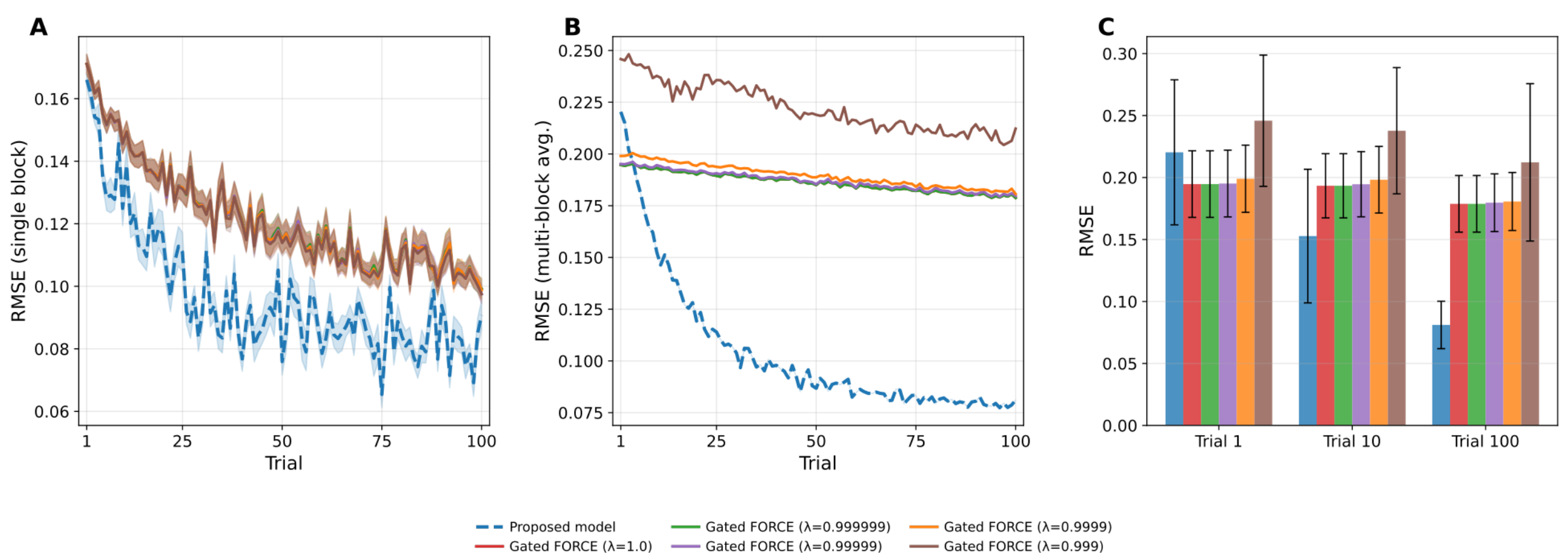

**A,** Root-mean-square error (RMSE) during single-block learning, plotted as a function of trial index (mean ± SEM across 20 simulations covering all condition combinations). Six models are compared: the Proposed model and five Gated FORCE variants with different forgetting factors ($\alpha$=1.0, 0.999999, 0.99999, 0.9999, 0.999). **B,** RMSE during multi-block adaptation (60 blocks), plotted as mean ± SEM across blocks for each model. **C,** RMSE at trials 1, 10, and 100 within blocks during multi-block adaptation. Bars show mean RMSE across 60 blocks with error bars indicating standard deviation.

## Supplementary Figure 2: Sensitivity of Gated FORCE to P matrix initialization parameters

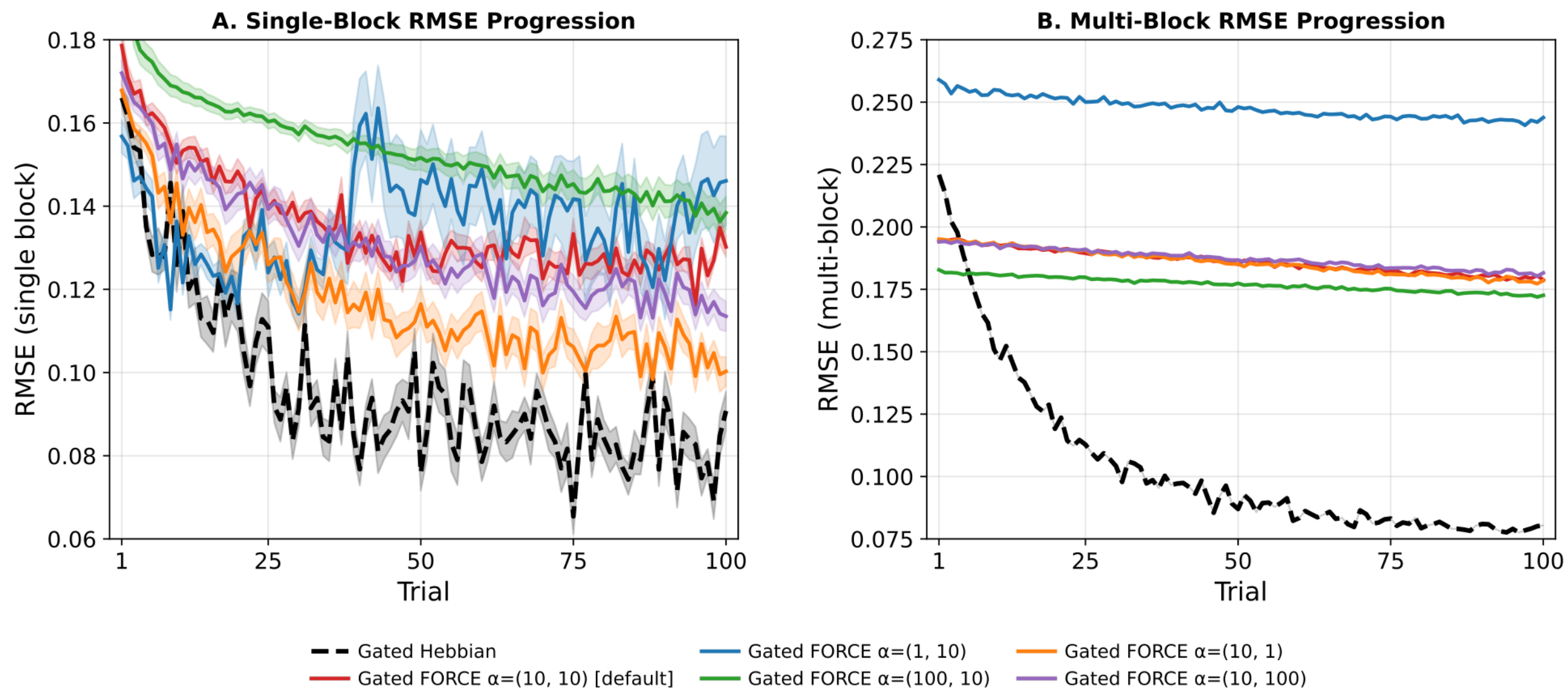

**A,** Root-mean-square error (RMSE) during single-block learning, plotted as a function of trial index (mean ± SEM across 20 simulations covering all condition combinations). Six models are compared: the Gated Hebbian model (proposed model, dashed black line) and five Gated FORCE variants with different P matrix initialization parameters (α_when, α_what), where the P matrix is initialized as $P_0 = I/\alpha$. The default configuration α=(10, 10) is shown in red; variants with faster temporal learning α=(1, 10), slower temporal learning α=(100, 10), faster identity learning α=(10, 1), and slower identity learning α=(10, 100) are shown in distinct colors. **B,** RMSE during multi-block adaptation (60 blocks with pseudo-randomized condition ordering), plotted as mean ± SEM across blocks for each model. Across all tested initialization parameters, the Gated FORCE model showed similar learning dynamics and did not match the performance of the proposed Gated Hebbian model, demonstrating that the performance difference between learning rules is robust to the choice of P matrix initialization scaling.

## Supplementary Table 1: Gated Local Hebbian Model design and architecture

| | Terminology | Value Used | Testable Range | Biological/Mechanical Relevance | Justification / Rationale | Ref |
|---|---|---|---|---|---|---|
| **A** | Reservoir size N | 1000 neurons | 500–5000 | Matches microcircuit scale; enough heterogeneity to span 10–100 ms windows | Balances capacity with efficiency; supports rich state dynamics without overfitting | 1,2 |
| **B** | Time step dt | 1 ms | 0.5–2 ms | Matches fast synaptic/axonal timescales; adequate for 10–100 ms timing | Resolves cue/teacher windows without unnecessary cost | 3 |
| **C** | Neuron model | Izhikevich (heterogeneous RS/FS) | RS/FS mixtures | Captures regular- (RS) and fast-spiking (FS) phenotypes; adaptation & resonance for timing | Biologically grounded dynamics with tractable cost | 4–6 |
| **D** | E/I ratio | 80% E / 20% I | 70/30–85/15 | Approximates cortical 4:1 excitatory–inhibitory composition | Supports balanced dynamics and stability | 7 |
| **E** | Recurrent weight scale g | 3.0 (scaled by $1/\sqrt{N}$) | 2.0–4.0 | Places network in rich, near-chaotic regime supportive of temporal basis | Yields diverse trajectories exploitable by linear readouts | 2,8 |
| **F** | Recurrent sparsity | 30% | 10–30% | Sparse long-range recurrence typical of cortex | Economical yet expressive connectivity | 7,9 |
| **G** | Dale's law in Ω and E | E→+ / I→− | Required | Preserves biological sign constraints on projections | Prevents unrealizable mixed signs per neuron | 7 |
| **H** | Synapse model | Double-exponential τr=2 ms, τd=20 ms | τr 1–3 ms; τd 10–50 ms | Approximates AMPA/GABA kinetics; realistic filtering | Maintains temporal precision with moderate integration | 10 |
| **I** | Input (cue) gain | 10 (a.u.) | 5–15 (a.u.) | Ensures cue perturbs reservoir without saturation | Drives state into predictive regime | 1,2 |
| **J** | Bias current $I_{bias}$ | Mean 1.0 (a.u.) (Gaussian) | 0.5–2.0 (a.u.) | Maintains baseline activity for eligibility accumulation | Prevents quiescence; supports sustained dynamics | 11 |
| **K** | Feedback matrix $E$ | Random, sign-constrained by E/I; gain 3.0 | 2–5 | Implements top-down/prediction feedback to reservoir | Enables on-line correction and credit signaling | 12,13 |
| **L** | Readout mask φ (fraction) | 30% of neurons | 10–40% | Sparse decoding consistent with selective downstream readouts | Reduces overfitting; improves generalization | 14 |
| **M** | Outputs | 2 channels (A,B) | Task-dependent | Implements 'what' readout and 'when' timing trace per channel | Minimal to demonstrate joint object | |
| **N** | 'when' learning rate η_when | 10 | 3–15 | Fast, online adaptation of latency predictions | Tracks nonstationary intervals within block | 15,16 |
| **O** | 'what' learning rate η_what | 2 | 1–15 | Slower, offline consolidation of identity probabilities | Stabilizes identity map under switching | 15,16,17 |
| **p** | Gate G(t) phases | Pre-detection (broad) → Post-detection (selective) | Binary/graded | Implements attention modulation of plasticity | Separates baseline shaping from credit assignment to active channel | 12,15,18,19 |
| **Q** | Three-factor plasticity for ‘when’ | $\Delta\varphi \propto$ presyn r × signed error × G(t) | Local only | Neuromodulated STDP/eligibility-trace style rule | Avoids global error broadcast/backprop | 15–17 |
| **R** | Trial structure | Jitter + 100ms pre cue + cue 30 ms + post cue 200 ms | Task-dependent | Spans subsecond regime relevant to cortical & cerebellar timing | Aligns with ITI/ISI used in interval tasks | 20,21 |
| **S** | ISI conditions (teacher windows) | 30-120 ms; 30 ms window | 10–200 ms | Creates separable latency bins for readout calibration | Tests multiplexed what × when learning | 20,21 |
| **T** | Jitter range | 0–200 ms | 0–250 ms | Prevents overfitting to exact onset; enforces temporal generalization | Builds hazard-like expectancy without clock | 22 |
| **U** | FORCE/RLS learning rate | η=10 (for when and what) | η=1–50 | Comparison to global-error methods | Benchmarks biological vs algorithmic tradeoffs | 23,24,25 |
| **V** | FORCE/RLS scaling before integration | s=50 | s=0.1-100 | Applies output scaling to $\mathbf{z}_{\text{what}}$ and $\mathbf{z}_{\text{when}}$ to have the network to be near an edge-of-rich-dynamics regime | Avoid overfit, divergence and improve stability | 23,24,25 |

**References used in Supplementary Table 1:**